\documentclass[12pt,twoside]{article}

\newcommand{\be}{\begin{equation}}
\newcommand{\ee}{\end{equation}}
\newcommand{\n}{\noindent}
\newcommand{\dd}{\,{\rm d}}

\newcommand{\ie}{{\it i.e.~}}

\usepackage{layout}
\usepackage{graphicx}
\usepackage[sort,authoryear,round]{natbib}
\usepackage{fancyhdr}
\usepackage{amssymb,amsfonts,amsmath}
\begin{document}
\n
{\large The Role of Lipid Bilayer Mechanics in Mechanosensation}\\

\n
{\normalsize
Tristan Ursell$^{1}$,
Rob Phillips$^{1,2,}$\footnote[1]{to whom correspondence should be sent:  phillips@pboc.caltech.edu},
Jan\'e Kondev$^{3}$,
Dan Reeves$^{3}$,
 Paul A. Wiggins$^{4}$
}\\

\n
\footnotesize $^{1}$Department of Applied Physics, California Institute of Technology, Pasadena CA 91125, USA\\
\footnotesize $^{2}$Kavli Nanoscience Institute, Pasadena  CA 91125, USA\\
\footnotesize $^{3}$Department of Physics, Brandeis University, Waltham  MA 02454, USA\\
\footnotesize $^{4}$Whitehead Institute, Cambridge MA 02142, USA\\

\normalsize\sloppy
\n
{\bf\small Abstract}\\
\n
Mechanosensation is a key part of the sensory repertoire of a vast array of different cells and organisms.   The molecular dissection of the origins of mechanosensation is rapidly advancing as a result of both structural and functional studies.  One intriguing  mode of mechanosensation  results from tension in the membrane of the cell (or vesicle) of interest. The aim of this review is to catalogue recent work that uses a mix of continuum and statistical mechanics to explore the role of the lipid bilayer in the function of mechanosensitive channels that respond to membrane tension.  The role of bilayer deformation will be explored in the context of the well known mechanosensitive channel MscL.  Additionally, we make suggestions for bridging gaps between our current theoretical understanding and common experimental techniques.\\

\n

\section{Mechanosensation and the Channels that Mediate it}

\n
Cells interact with each other and with their external environment.   These interactions are enabled by transmembrane proteins -  machines that have evolved to allow cells to detect and respond to changes in their environment.  These proteins detect external cues, such as an increase in ligand concentration or the presence of forces or voltage, and transiently alter the permeability of the cell membrane allowing ions, water, or even larger molecules to cross as well as triggering receptors for signaling  \citep{Clapham2001,Barry2005}.  The passage of these ions (or molecules) and the triggering of receptors then leads to a series of downstream events within the cell, enabling a response to these environmental cues. 

Mechanical forces and their corresponding deformations constitute one of the most important classes of external cues. Mechanosensation is a widespread phenomenon in a host of different single-celled and multicellular organisms \citep{Sachs1991,Kloda2001,Gillespie, Fain,Katsumi2004,Nauli2004}.    In bacteria, experimental evidence suggests that mechanosensation arises to detect and regulate the response to changes in the osmotic environment \citep{SukharevAnnRev1997,Rees1998,Pivetti2003}. More generally, the issue of cell shape and its attendant deformation is important not only in the context of osmotic stress and the management of physical stresses to which membranes are subjected \citep{Morris2001}, but also arises in context of remodeling of the cell and organelle membranes during cell division \citep{Kamada1995,Strange2001}.

In multicellular organisms, mechanosensation is important in a variety of ways.  One intriguing class of mechanosensors is linked to motility.  For example,  in nematodes like the much studied {\it C. elegans}, mechanosensation permits the worm to decide which way to move and may have a role in detecting body curvature, thus telling the worm when to change its wave-like shape \citep{Gillespie}.  Similarly, flies have hair bristles that respond to touch \citep{Duggan2000}, while the mechanosensitive lateral-line organelles in zebrafish provide the means for detecting directional water movement in a way very similar to the workings of our inner ear \citep{Gillespie}.  In each of these cases, genetics has led to the identification of a variety of genes implicated in the ability of the organism to respond to some form of mechanical stimulus. Parallel insights have been obtained in plants ({\it Arabidopsis} in particular), with the identification of a collection of novel proteins
that also appear to be mechanosensitive \citep{Haswell}.

Mechanosensitive ion channels are a class of membrane proteins that have recently garnered significant interest.  Genetic, biochemical and structural studies all conspire to make this a particularly opportune time to demand a more quantitative picture of the function of these channels.    In particular, there is a growing list of success stories in which the structures of channels associated with mechanosensation  have been found in both closed and open states \citep{Rees1998,Bass2002, PerozoNature2002, PerozoCOSB2003}.  In addition, functional studies that probe how gating depends upon membrane tension or external forces are beginning to make it possible to dissect the various contributions to the energetics of channel gating \citep{SukharevJGP1999,PerozoNSB2002,SukharevBJ2004,Akitake2005}.

As a result of these studies, a number of ideas have been proposed to explain the different ways in which external force can couple to membrane-protein conformation. Two modes of action that have been hypothesized for channels are: i) cases in which physical, polypeptide linkers pull on some part of the protein resulting in gating, ii) cases in which tension in the surrounding bilayer forces the channel to open. The aim of this article is to show how statistical mechanics and simple models of bilayer elasticity can be used to glean insights into this second class of mechanosensors.

The remainder of the article is built in four main sections.  In the next section, we describe how statistical mechanics can be used to analyze the probability that a two-state mechanosensitive channel is open.  This discussion will include an analysis of how the external load (\ie  the tension) can be included in the statistical mechanical treatment of these problems.  The next section
considers the elastic deformations imposed on a bilayer by the presence of a transmembrane protein, and shows how these deformations result in a mechanosensitive channel acting as a bistable switch (\ie a protein with two stable conformations). In the subsequent section, we discuss experimental considerations that will help form a tighter connection between theory and experimental techniques.  Finally, we examine  the way multiple channels in a membrane might interact through the intervening lipid bilayer and how these interactions can alter the conformational statistics of individual channels. \\

\section{Statistical Mechanics of Mechanosensitive Channels}

To begin, we review the application of statistical mechanics to a simple two-state mechanosensitive channel. This analysis will serve as the starting point for our subsequent, more detailed analysis which explores how bilayer elasticity can contribute to the energetics of the closed and open states of a channel.

\subsection{Lipid Bilayer vs. Protein Internal Degrees of Freedom}

One convenient scheme for characterizing the state of ion channels is to invoke the state variable $\sigma$, which is defined by $\sigma=0$ if the channel is closed and $\sigma=1$ if the channel is open.    Our aim is to compute the open probability $P_{open}$ which, in terms of our state variable $\sigma$, can be written as $\langle \sigma \rangle$, where $\langle \cdots \rangle$ denotes an average. When $\langle \sigma \rangle \approx 0$, this means that the probability of finding the channel open is low.  Similarly, when $\langle \sigma \rangle \approx 1$, this means that it is almost certain that we will find the channel open. To evaluate these probabilities we need to invoke the Boltzmann distribution, which tells us that the probability of finding the system in a state with energy $E(\sigma)$ is $p(\sigma)=e^{-\beta E(\sigma)}/Z$, where $Z$ is the partition function defined by $Z=\sum_{\sigma} e^{-\beta E(\sigma)}$, $\beta=1/k_BT$, $k_B$ is Boltzmann's constant, and $T$ is the temperature in degrees Kelvin.

\begin{figure}
\begin{center}
%\setbmp{-13cm}{0cm}{5cm}{channelstatesweights.jpg}
\includegraphics[width=13cm]{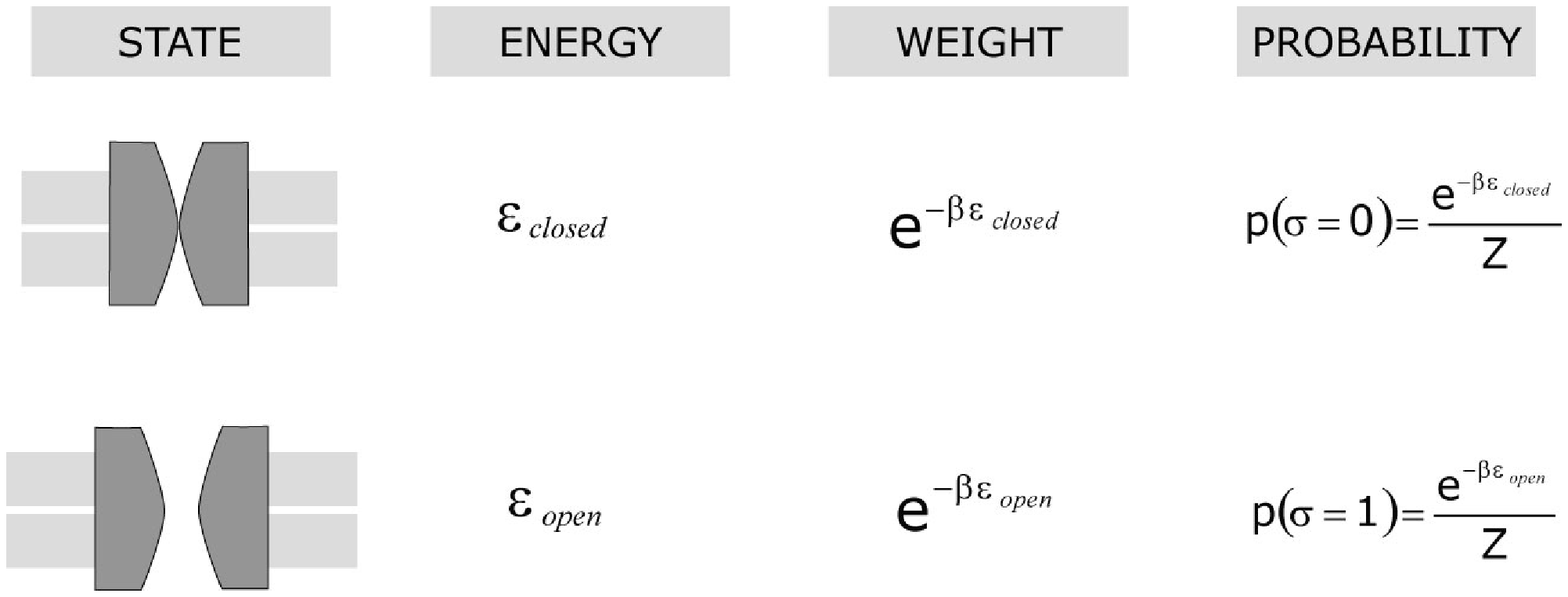} 
\caption{States, Boltzmann weights and corresponding probabilities for a two-state  ion channel.   The two different states have different energies and the probability of these different states is determined by the Boltzmann distribution. }
\label{ChannelStatesWeights}
\end{center}
\end{figure}

On the level of a single channel, we introduce  $\epsilon_{closed}$ and $\epsilon_{open}$ for the energies of the closed and open states, respectively, as shown in Figure \ref{ChannelStatesWeights}.  These energies contain contributions from deformations of the surrounding lipid  bilayer as well as internal protein energetics; however, they do not contain the tension-dependent driving force which we will address separately.  The state variables can be used to write the channel energy (in the absence of tension) as
\begin{equation}
E(\sigma)=(1-\sigma)\epsilon_{closed}+\sigma \epsilon_{open}.
\end{equation}  
With these energies in hand, we can assign weights to the different states as shown schematically in  Figure \ref{ChannelStatesWeights}. Within this scheme, the probability that the channel is open is given by $\langle \sigma \rangle$ and can be computed as $\langle \sigma \rangle =\sum_{\sigma=0}^1 \sigma p(\sigma)$, where $p(\sigma)$ is the probability of finding the channel in state $\sigma$. To compute these probabilities, we invoke the Boltzmann distribution, and evaluate the partition function given by 
\begin{equation}
Z=\sum_{\sigma=0}^1 e^{-\beta E(\sigma)}=e^{-\beta \epsilon_{closed}}+e^{-\beta \epsilon_{open}}.
\end{equation}
As a result, we see that the open probability can be written as
\begin{equation}
\langle \sigma \rangle ={e^{-\beta \epsilon_{open}} \over e^{-\beta \epsilon_{closed}}+e^{-\beta \epsilon_{open}}}=\frac{1}{1+e^{\beta(\epsilon_{open}-\epsilon_{closed})}}.
\end{equation}
This expression is relatively sterile in the absence of some term that tunes the energies of the open and closed states to reflect the impact of external driving forces.  In fact, one of the most remarkable features of ion channels is that the probability of being in different states can be tuned by external factors such as ligand concentration, the application of a voltage, or application of tension in the surrounding membrane. In general, this formalism can account for any of these driving forces, but we will restrict our attention to the important case of mechanosensitive channels, where the key driving forces are mechanical.  In this case,  gating occurs  when the energy balance between  the open and closed states is altered by membrane tension. 
 
\begin{figure}
\begin{center}
%\setbmp{-7cm}{0cm}{7cm}{loadingdevice.jpg}
\includegraphics[width=6cm]{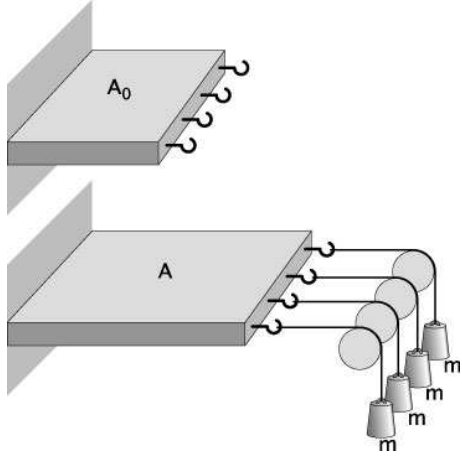}
\caption{Energy of the loading device for membrane deformation.  This figure compares the unloaded and loaded membrane and shows how membrane deformation results in a {\it lowering} of the potential energy of the loading device.  In this hypothetical experiment, the tension (force per unit edge length) in the membrane is given by $\tau=mg / \Delta l$ where $\Delta l$ is the distance between two consecutive hooks, and $g$ is the acceleration due to gravity.}
\label{PELoadingDevice}
\end{center}
\end{figure}

To give the origin of membrane tension a physical meaning, we introduce the notion of a ``loading device'', which we define as the external agent acting on a lipid bilayer to alter its tension. As depicted by hanging weights on the bilayer in Figure \ref{PELoadingDevice},  we can make a toy model of how changes in bilayer geometry are coupled to the energy of this loading device.  The point of introducing this hypothetical situation is to enforce the idea that, in our statistical mechanical treatment of this problem, the loading device is an important part of the overall free energy budget of the system.  As a result, when we write down the partition function for a problem involving a channel and a deformable membrane, we have to account for the internal protein energetics, the deformation energy of the lipid bilayer, and the energy associated with the loading device itself.  In particular, we note that an increase in the membrane area will lead to a lowering of the weights depicted in Figure \ref{PELoadingDevice} and a corresponding decrease in the energy of the loading device.  Of course, the application of tension in real membranes is not performed by hanging weights, but through techniques such as micropipette aspiration \citep{GoulianBJ1998,Evans2000}.  Nevertheless, the concept of hanging weights brings the importance of the energy of the loading device into sharp focus.

\begin{figure}
\begin{center}
%\setbmp{-12cm}{0cm}{4cm}{channelload.jpg}
\includegraphics[width=13cm]{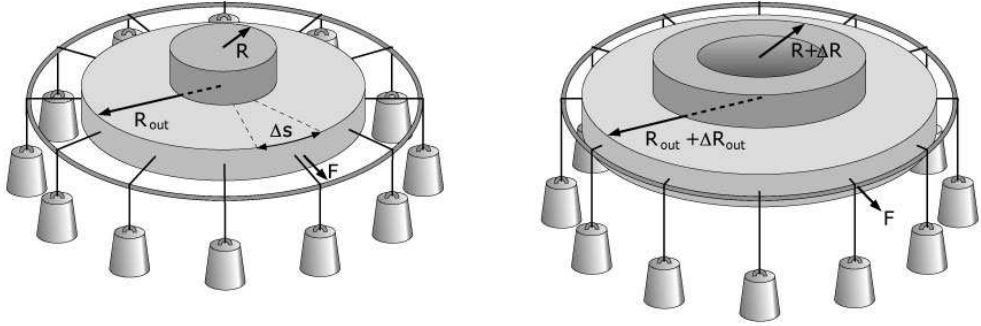}
\caption{Schematic of how channel opening results in a relaxation in the loading device.   For simplicity, we represent the loading device as a set of weights attached to the membrane far from the channel.  When the channel opens, these weights are lowered, and the potential energy of the loading device is decreased.}
 \label{MscL2}
\end{center}
\end{figure}

For the case of tension-activated ion channels, the open probability, $\langle \sigma \rangle$, is dictated by a competition between the energetic advantage associated with reduction in the energy of the loading device and the energetic cost of the open state due to both the internal protein energetics and the energetics of membrane deformation.    Following up on the idea of Figure \ref{PELoadingDevice}, but now with special reference to the case of a mechanosensitive ion channel, Figure \ref{MscL2} shows how the opening of the channel results in a reduction of the energy of the loading device.

The total area of the bilayer is constant (to within a few percent), and as a result, when the channel opens and the radius gets larger the weights in our hypothetical loading device  are lowered by some amount, which lowers the potential energy.   The greater the weights, the larger the change in potential energy.    The notion of weights is a simple representation of externally applied forces on the membrane. If we imagine a finite membrane with fixed area as shown in Figure \ref{MscL2}, when the channel opens, the outer radius will change as $\Delta R_{out}=(R/R_{out})\Delta R$, where $R$ is the closed channel radius, $\Delta R$ is the change in channel radius upon opening, $R_{out}$ is the outer radius of the membrane when the channel is closed, and $\Delta R_{out}$ is the increase in the outer radius of the membrane when the channel opens.   We are interested in evaluating the change in potential energy of the loading device (\ie  the dropping of the weights) as a result of channel opening.  To do so, we compute the work associated with the force $F$, which is most conveniently parameterized through a force per unit length (the tension, $\tau$) acting through the distance $\Delta R_{out}$ as shown in Figure \ref{MscL2}.  This results in
\begin{equation}
\Delta G_{tension}=\underbrace{\tau \Delta s}_{\mbox{\footnotesize force on arc}} \times \underbrace{\frac{R}{R_{out}}\Delta R}_{\mbox{\footnotesize displacement of arc}} \times \underbrace{\frac{2\pi R_{out}}{\Delta s}}_{\mbox{\footnotesize number of arcs}}.
\end{equation}
where $\Delta G$ represents a change in free energy.  We have introduced the variable $\Delta s$ for the increment of arc length such that $\tau=F/\Delta s$.   Given these definitions, we see that the change in the energy of the loading device is given by
\begin{equation}
\Delta G_{tension} = -\tau 2\pi R \Delta R.
\end{equation}

In light of our insights into the energy of the loading device, we introduce the energy as a function of the applied tension $\tau$, which is given by
\begin{equation}
E(\sigma,\tau)=(1-\sigma) \epsilon_{closed}+\sigma \epsilon_{open}-\sigma \tau \Delta A.
\label{IonChannelEnergy}
\end{equation}
The term $-\sigma \tau \Delta A$ favors the open state and reflects the fact that the energy of the loading device is lowered in the open state.    In fact, this term reveals that any increase in protein area is energetically favored when membrane tension is present, which could imply hidden mechanosensitivity in other classes of ion channels and receptors - a subject we will touch upon later in this review.

\begin{figure}
\begin{center}
%\setbmp{-13cm}{0cm}{7cm}{statesweightsload.jpg}
\includegraphics[width=13cm]{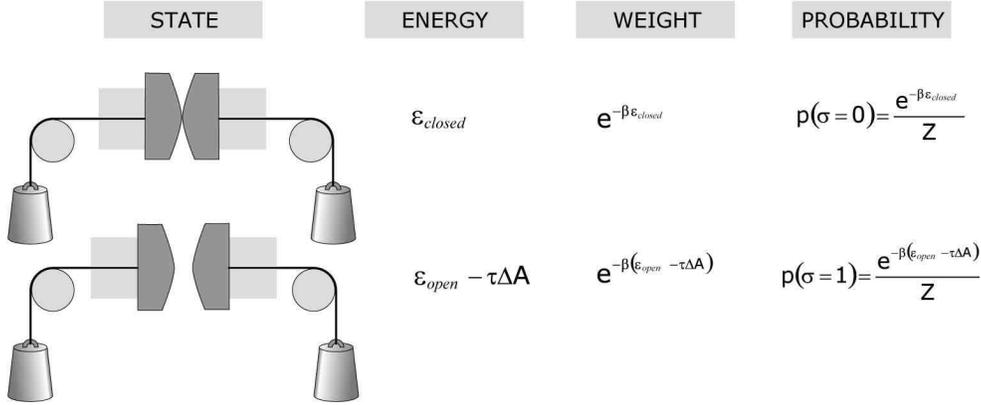}
\caption{States, weights and corresponding probabilities for a two-state mechanosensitive channel under load.}
\label{StatesWeightsLoad}
\end{center}
\end{figure}

\begin{figure}
\begin{center}
%\setbmp{-9cm}{0cm}{5cm}{channelopenprobtension.jpg}
\includegraphics[width=8cm]{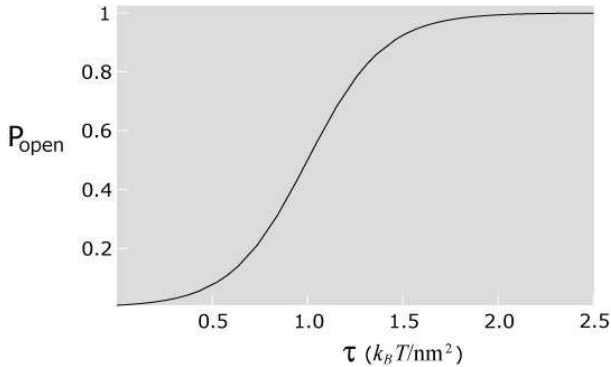} 
\caption[]{Ion channel open probability as a function of applied tension. The plot shows $P_{open}=\langle \sigma \rangle$ as a function of the applied tension $\tau$.  The parameters used in the plot for a model mechanosensitive channel are $\epsilon_{open}-\epsilon_{closed}=10\,k_BT$ and $\Delta A=10\,\mbox{nm}^2$.  The critical tension is $1.0\,k_BT/\mbox{nm}^2$, corresponding to $P_{open}=1/2$ .  For reference, the tension can be rewritten as 1 pN/nm $\simeq0.25\,k_BT/\mbox{nm}^2$.}
\label{IsingChannel}
\end{center}
\end{figure}

To compute the open probability of the channel in the presence of applied tension, we need to once again evaluate the partition function $Z=\sum_{\sigma} e^{-\beta E(\sigma)}$. Using the energy given in eqn.~\ref{IonChannelEnergy}, we find
\begin{equation}
Z=e^{-\beta \epsilon_{closed}}+e^{-\beta(\epsilon_{open} - \tau \Delta A )}.
\end{equation}
This permits us to write down the open probability directly as 
\begin{equation}
P_{open}={e^{-\beta (\epsilon_{open}-\tau \Delta A)} \over e^{-\beta (\epsilon_{open}-\tau \Delta A)}+e^{-\beta \epsilon_{closed}}}=\frac{1}{1+e^{\beta(\epsilon_{open}-\epsilon_{closed}-\tau\Delta A)}}.
\end{equation}
The corresponding states, weights, and probabilities for a channel under applied tension are shown in Figure \ref{StatesWeightsLoad}.  The open probability of a mechanosensitive channel is shown in  Figure \ref{IsingChannel} as an increasing function of the applied tension.

To understand how a particular channel is going to behave under a driving force, we need to know two things.  First, we need to understand the channel's intrinsic preference for each of its two states, which is encoded by $\epsilon_{closed}$ and $\epsilon_{open}$.  Second, we need to understand how the external driving force alters the relative energies of these different states.    With these two quantitative measurements in hand, statistical mechanics allows us to compute the behavior of the channel under a range of driving forces. To make further progress, we need to examine the microscopic origins of $\epsilon_{closed}$ and $\epsilon_{open}$.    Intriguing recent experiments suggest that these energies are driven in large measure by membrane deformations.

\section{Bilayer Free Energy and Gating of a Mechanosensitive Channel} 

\begin{figure}
\begin{center}
%\setbmp{-8cm}{0cm}{6cm}{perozo.jpg}
\includegraphics[width=8cm]{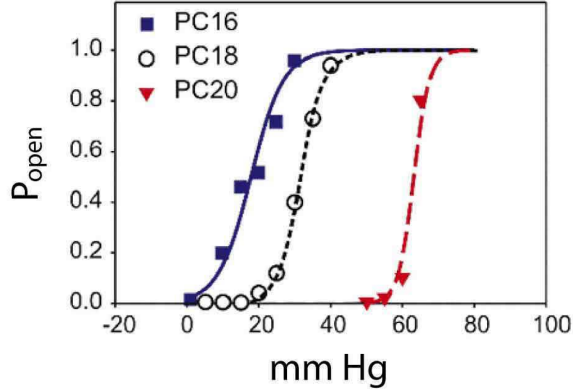} 
\caption{Ion channel open probability for different lipids.  The graph shows the open probability of the mechanosensitive channel MscL as a function of the applied pressure across the bilayer for three different lipid tail lengths.  Pressure difference is related to bilayer tension via a constant, and hence this suggests that bilayer thickness (for carbon tail lengths of 16, 18 and 20) affects channel function. Figure adapted from \protect\citet{PerozoNSB2002}.}
\label{ChannelData}
\end{center}
\end{figure}

The abstract formalism of the previous section leaves us poised to examine mechanosensation to the extent that we can understand the physical origins of $\epsilon_{closed}$ and $\epsilon_{open}$.   The main idea of this part of the review is to show how simple models of the elastic properties of lipid bilayers can be used to determine the bilayer's contribution to $\epsilon_{closed}$ and $\epsilon_{open}$. One of the key clues that hints at the importance of membrane deformation in dictating channel gating is the data shown in Figure \ref{ChannelData}.   In particular, this plot shows how the open probability depends upon the lipid carbon tail length.    This data strongly suggests that the energetics of the surrounding membrane is an important part of the overall free energy budget of channel gating (also see \citet{Martinac2002} and the informative review by \citet{Jensen2004}).

The parameters $\epsilon_{closed}$ and $\epsilon_{open}$ can each depend on some combination of the energetics of protein conformation, membrane deformation, and hydration energy. Our strategy is to use the tools of continuum mechanics to calculate how the deformation of lipids surrounding a protein and the applied tension work in concert to affect the channel's preference for a particular state \citep{Huang1986,  Dan1998, AndersenBJ1998,Wiggins2005}.  Unfortunately, relatively little is known about how the internal rearrangements of the protein and the hydration energy of the channel pore contribute to the overall free energy balance \citep{Yoshimura1999, Anishkin2005}.  This ignorance is in part due to a lack of general rules that tell us how internal rearrangements translate into changes in protein energy. Further, the lack of  crystal structures in the open and closed states of many channel proteins means we cannot be sure where each residue moves, which are exposed to the surrounding lipids and which are facing the hydrated internal pore.  It is also difficult for molecular dynamics to comment on the energies associated with the internal movements of the protein \citep{Gullingsrud2001, Gullingsrud2003, Elmore2001,Elmore2003} because the all-atom energies of these simulations are very large in comparison to the changes in free energy, and hence it is difficult to distill relatively small free energy changes in the background of large energy fluctuations.  To complicate the issue further, it is also possible that the internal movements of the protein yield relatively small free energy changes between the two states,  but may provide various kinetic hurdles in the form of energy barriers, which affect the transition {\it rate} from one state to another. 

It is reasonable on the scale of a single membrane protein to ask whether a bilayer composed of discrete lipid molecules can be approximated as a continuum material.  We argue heuristically that, given the relative diffusion coefficients of membrane proteins ($D\sim0.1-1\mu\mbox{m}^2/s$) \citep{Doeven2005,Gambin2006,Guigas2006} and lipids ($D\sim10\mu\mbox{m}^2/s$) \citep{Kahya2003}, in the time it takes a transmembrane protein to diffuse one lipid diameter, many lipids will have exchanged places near the protein, in a sense averaging out the discreteness of the lipid molecules.  Additionally, the transition time for protein conformational change ($\sim5\,\mu s$) \citep{Lester2004} is slow compared to lipid diffusion.  Hence, we argue the bilayer can be approximated as a continuous material in equilibrium with well-defined elastic properties \citep{HuangBJ1999}.  Further, we choose to formulate our analysis in the language of continuum mechanics, rather than  lateral pressure profiles \citep{Cantor1999}.  

Approximating the membrane as a continuum material  \citep{Helfrich1973,Huang1986,Dan1998,AndersenBJ1998,HuangBJ1999,Wiggins2004,Wiggins2005}, we will concentrate our analysis on how the mechanical properties and deformations of lipids affect the energy balance of the protein, and how tension can play the role of a driving force for gating the channel.  In particular, the mechanosensitive channel of large conductance (MscL) is one of the best characterized mechanosensitive channels.  Additionally, a combination of X-ray crystallography and electron paramagnetic resonance studies have yielded insights into the structures of both the closed and open states of MscL \citep{Rees1998,PerozoJGP2001, PerozoNature2002}.  One of the outcomes of this structural analysis is the idea that the structure can be roughly approximated as a cylinder, making  it amenable to mechanical modeling.  MscL exemplifies many of the characteristics one might call ``design principles'' for a mechanosensitive channel \citep{Wiggins2004, Wiggins2005}, such as change in hydrophobic thickness, a change in radius, and sensitivity to membrane curvature.  In the remainder of the review, we will lay the foundation for a continuum mechanical understanding of how lipid deformations and tension work together to give a switchable channel.

\subsection{The Case Study of MscL}

In the prokaryotic setting, the physiological purpose of MscL is thought to be an emergency relief valve under conditions of hypoosmotic shock \citep{SukharevAnnRev1997,Rees1998,Pivetti2003}, whereby the osmotic pressure difference between the inside of a cell and the environment translates into increased membrane tension.  The channel responds by gating and non-selectively releasing osmolytes to the environment until the internal and external pressures are equilibrated \citep{SukharevJGP1999,SukharevNature2001}.  This presents us with (at least) two key questions.  First, what gives MscL its ability to ``sense'' tension in the membrane?  Second, what role is the lipid bilayer playing in the gating transition? 

We will argue that the answers to these questions are found in the  properties of a lipid bilayer and the geometrical features of the channel as revealed in  Table~\ref{tb1}.  In particular, the bilayer has four key elastic properties that give it the ability to transduce tension and resist deformation by a transmembrane protein.   The most striking elastic feature is the in-plane fluidity of the bilayer, which, in the absence of cytoskeletal interactions, results in  equalization of tension throughout the membrane.  This means that any in-plane stress (\ie tension) on the membrane is felt everywhere equally.  Hence, in the case of MscL, an increase in tension is applied uniformly to the outer edge of the protein, essentially trying to ``pull''  the channel open.  We argue it is this ``pulling'' which constitutes the driving force for channel gating.  However, this driving force is competing with the energetic cost to gate the channel due to internal conformational changes within the protein and deformations of the surrounding lipid.  

Three other properties give the membrane the ability to store energy elastically upon deformation.  First, each leaflet of the membrane resists changes in the angle between adjacent lipid molecules, leading to bending stiffness of the membrane \citep{Helfrich1973,Huang1986, Dan1998, AndersenBJ1998,HuangBJ1999, Wiggins2005}.  Second, the membrane has a preferred spacing of the lipid molecules in-plane and will resist any changes in this spacing due to external tension \citep{Dan1998,Evans2000}. Third, the membrane has a well-defined equilibrium hydrophobic thickness which, when given an embedded protein of a different hydrophobic thickness, leads to energetically costly `hydrophobic mismatch' \citep{Dan1998,AndersenBJ1998,HuangBJ1999,Wiggins2004,Wiggins2005}.

The competition between the driving force and the energetic cost to gate the channel hints at a set of design principles that dictate how the channel behaves as a bistable switch.  If we neglect the molecular details of MscL, its conformational change can be characterized by a set of simple changes in geometrical  parameters.  In particular, in our coarse-grained description we will think of the  gating transition as being accompanied by changes in height, radius and protein angle, all of which couple to various modes of membrane deformation as shown in Figure \ref{defmodes}.  The central question becomes, is deformation of the lipids surrounding the protein a major player in gating energetics?  Indeed, experiments have already suggested that the gating characteristics are intimately linked to the hydrophobic mismatch between the protein and bilayer as was shown in Figure \ref{ChannelData}  \citep{PerozoNSB2002,Martinac2002,Jensen2004}.  It is the goal of the following sections to build up a theoretical framework for understanding the various kinds of bilayer deformation around a transmembrane protein and to describe how these deformations contribute to the overall free energy budget associated with the gating of MscL (and probably other channels as well). 

\n
\begin{center}
\begin{table}
\begin{center}
\begin{tabular}{|l|l|l|}
\hline
Parameter: & Value: & Source:\\
\hline
Closed height	&	3.8 nm	&	\citet{Rees1998}\\
\hline
Closed radius	&	2.5 nm	&	\citet{Rees1998}\\
\hline
Open height	&	2.5 nm	&	\citet{PerozoNature2002}\\
\hline
Open radius	&	3.5 nm	&	\citet{PerozoNature2002}\\
\hline
Measured $\Delta A^{*}$	&	$20\,\mbox{nm}^2$	&	\citet{SukharevBJ2004}\\
\hline
Measured $\Delta G^{*}$	&	$51\,k_BT$	&	\citet{SukharevBJ2004}\\
\hline
Calculated $\Delta G^{*}$ (at critical tension)	&	$\sim55\,k_BT$	&	this article\\
\hline
Critical Tension$^{*}$ 	&	$\sim2.5\,k_BT/\mbox{nm}^2$		&	\citet{SukharevBJ2004}\\
\hline
Lytic Tension$^{*}$		&	$\sim3.5\,k_BT/\mbox{nm}^2$	&	\citet{Evans2000}\\
\hline
Bending Modulus ($\kappa_b$)		&	$\sim20\,k_BT$	&	\citet{Helfrich1995}\\ 
											&		&	 \citet{Evans2000}\\
\hline
Area Stretch Modulus ($K_A$)		&	$\sim60\,k_BT/\mbox{nm}^2$	&	\citet{Evans2000}\\
\hline
Leaflet Thickness ($l$)		&	$1.75\,\mbox{nm}$	&	\citet{Evans2000}\\
\hline
\end{tabular}
\caption{MscL geometrical and bilayer elastic parameters. $^{(*)}$ These parameters depend on the elastic properties of the bilayer, in particular the bilayer bending modulus ($\kappa_b$), the bilayer area stretch modulus ($K_A$), and the leaflet hydrophobic thickness ($l$).
\label{tb1}}
\end{center}
\end{table}
\end{center}

\subsection{Bilayer Deformation, Free Energy and the Role of Tension}

To investigate the contribution of membrane deformation to channel gating in mechanosensitive channels, we put  our ignorance of the internal protein energetics aside and focus on the response of the membrane.   The point of this analysis is to see how large the membrane contributions are to the free energy of channel gating, and to examine how they compare to the measured values.   A mechanosensitive channel must resist the driving force due to tension to exhibit the properties of a bistable switch.   As we will demonstrate in this section, deformation of the surrounding lipids can provide this resistance, and almost certainly does in the case of MscL, given our knowledge of the open and closed structures and the body of experimental data describing the interactions between lipids and MscL \citep{SukharevJGP1999,PerozoNSB2002,Powl2003}.  

The deformations that a transmembrane protein induces can be most broadly split into two main classes:  those that deform the midplane of the bilayer, and those that deform the bilayer leaflet thickness.  If the deformation is not too severe, these two types of deformation are independent of one another \citep{Wiggins2005}.   Figure \ref{defmodes} shows these two classes of deformation and the simple model idealizations implied by elastic descriptions.    The basic structure of the models we consider are those in which the contributions of deformation to the overall free energy are obtained by computing local bending and thickness deformation, and then summing over the contributions from all the area elements making up the bilayer.  

\subsubsection{Midplane Deformation}
\label{middef}

\begin{figure}
\begin{center}
%\setbmp{-13cm}{0cm}{6cm}{defmodes.jpg}
\includegraphics[width=13cm]{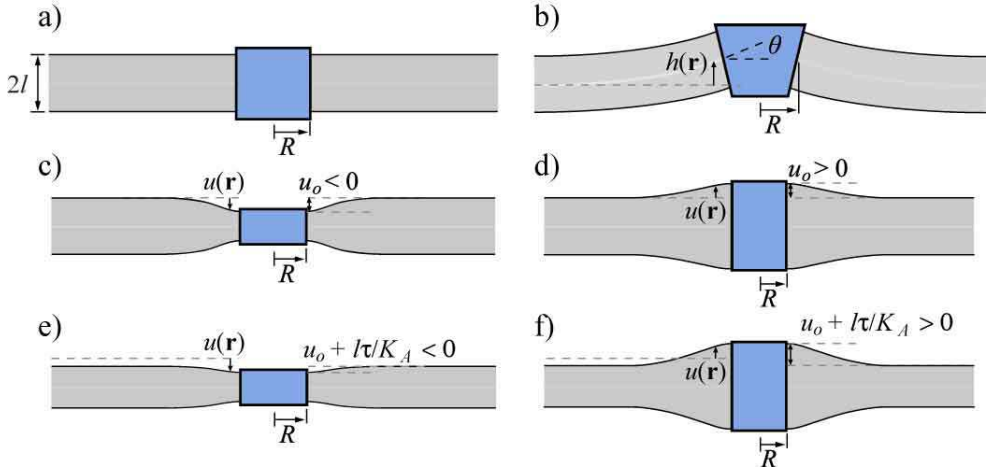}
\caption{Modes of bilayer deformation.  Differences in the equilibrium shape of the membrane and an embedded protein give rise to local deformations.  a) The  undeformed  state is a transmembrane protein with zero hydrophobic mismatch and a flat midplane. Even an initially undeformed membrane exhibits tension dependence since tension induces bilayer thinning.  b) An angled protein induces midplane bending, characterized by the function $h({\bf r})$ and the boundary slope $\theta$.  As tension increases, the most preferred energetic state is $\theta=0$. c)  A membrane protein that is thinner than the equilibrium thickness of the bilayer compresses the bilayer causing local area expansion and bending of each leaflet, characterized by the function $u({\bf r})$.  d) A membrane protein that is thicker than the equilibrium thickness of the bilayer stretches the bilayer causing local area reduction and bending of each leaflet.  e) An increase in tension will decrease the energetic cost of a membrane protein that is thinner than the equilibrium thickness of the bilayer, as the membrane thins and approaches zero hydrophobic mismatch. f) Likewise, an increase in tension will increase the energetic cost of a membrane protein that is thicker than the equilibrium thickness of the bilayer.}
\label{defmodes}
\end{center}
\end{figure}

Deformation of the midplane of the bilayer involves a cost to bend the midplane from its flat, equilibrium position \citep{Helfrich1973, HuangBJ1999,Wiggins2005}.  We use the function $h({\bf r})$ to denote this change in height of the bilayer midplane as a function of the position ${\bf r}$ as shown in Figure \ref{defmodes}.  The energy cost associated with bending the membrane away from its flat configuration can be written as
\be{
G^{(\mbox{\tiny mid})}_{\mbox{\tiny bend}}=\frac{\kappa_b}{2}\int\left(\nabla^2 h({\bf r})-C_o\right)^2\dd^2{\bf r},
\label{Gmidfunct}
}\ee
where the bilayer bending modulus $\kappa_b\simeq20\,k_BT$ \citep{Helfrich1995,Evans2000} and $C_o$ is the midplane spontaneous curvature.  Throughout the review the gradient operator is defined by $\nabla = \left(\partial/\partial x, \partial/\partial y\right)$, and the Laplacian operator by $\nabla^2 = \partial^2/\partial x^2+\partial^2/\partial y^2$, in Cartesian coordinates.   In general, bilayers with symmetric leaflet compositions have zero midplane spontaneous curvature. Tension also plays a role in the energetics of midplane deformation because any bend in the midplane results in a reduction in the projected area of the membrane, which couples directly to an increase in the energy of the loading device.  This effect is quite intuitive when one considers deformations of a macroscopic membrane under tension and results in a contribution to the free energy of the form 
\be{
G^{(\mbox{\tiny mid})}_{\mbox{\tiny ten}}=\frac{\tau}{2}\int(\nabla h({\bf r}))^2\dd^2{\bf r},
}\ee
where the tension, $\tau$, ranges from zero up to the nominal membrane lytic tension of $\sim3.5\,k_BT/\mbox{nm}^2$ \footnote[1]{The lytic tension of a bilayer is technically a dynamic quantity \citep{Evans2003}, however, we quote the lytic tension as the tension at which bilayer lysis is a rapid, spontaneous process.} \citep{Evans2000}.  In general, the elastic parameters we use are representative of a typical phosphatidylcholine (PC) lipid. Thus the total energy expended to deform the midplane over an area $A$ is 
\be{
G^{(\mbox{\tiny mid})}=\int_{A}\left(\frac{\tau}{2}(\nabla h({\bf r}))^2+\frac{\kappa_b}{2}(\nabla^2 h({\bf r}))^2\right)\dd^2{\bf r}.
\label{gmid}
}\ee
The logic behind this kind of analysis is to find the free energy minimizing function $h({\bf r})$. One way to carry out this minimization is by solving a partial differential equation that is generated by formally minimizing the free energy.  An alternative (and approximate) scheme to be explored later in this section is to make a guess for the functional form of $h({\bf r})$ and to minimize with respect to some small set of parameters.  This approach is called a variational method and can be quite useful for developing intuition.

In the midplane-deforming model, the protein can dictate the slope of the membrane at the protein-lipid interface which, in addition to the protein radius, will determine the deformation energy.  The length scale over which the membrane returns to its unperturbed state is given by $\sqrt{\kappa_b/\tau}$ and the energy for this type of deformation is
\be{
G^{(\mbox{\tiny mid})}(R,\tau)=\theta^2\pi R\sqrt{\kappa_b\tau}\frac{K_0(R\sqrt{\tau/\kappa_b})}{K_1(R\sqrt{\tau/\kappa_b})},
\label{gmidenergy}
}\ee
where $R$ is the radius of the protein, $\theta$ is the slope of the membrane at the protein-lipid interface as shown in Figure \ref{defmodes}, and $K_i$ are modified Bessel functions of the second kind of order $i$ \citep{Turner2004, Wiggins2005}.  Given a protein with a particular radius and fixed boundary slope, an increase in tension will make any deformation {\it more} costly.  Hence, for midplane deformation, increased tension prefers a flatter membrane and/or smaller protein radius.  To get a feel for the energy scale of this deformation several examples for different parameter values are summarized in Table \ref{tb2}. 

With the contribution to the free energy difference arising from midplane deformation in hand, we can now explore the competition between applied tension and the energetics of membrane deformation in dictating channel gating. The key to understanding the interplay between tension and deformation energetics lies in the scaling of these two effects with protein radius.  The midplane deformation energy scales roughly linearly with the radius of the protein and is unfavorable.  On the other hand, the term proportional to the applied tension  scales as the square of the protein radius and favors the open state. If we fix the membrane slope, then the energy of a midplane deforming protein as a function of protein radius and tension is
\be{
G(R,\tau)\simeq\underbrace{G^{(\mbox{\tiny mid})}(R,\tau)}_{\mbox{\footnotesize membrane}}-\underbrace{\tau\pi R^2}_{\mbox{\footnotesize loading device}}.
\label{midnucleation}
}\ee 
%We note that at zero or very low tensions, this energy imparts no radial preference, an undesirable quality for a bistable switch. However, for low but nonzero tension the midplane deformation dominates, and the net result is a preference for a smaller protein radius.  
As tension increases, the potential energy of the loading device will eventually overcome the deformation energy and a larger protein radius will be the preferred state.  Indeed, midplane deformations have been hypothesized to be an important functional mechanism of MscL \citep{Turner2004}.   One of the uncertainties that accompanies a model of this type is the fact that there is some function that connects the slope of the membrane at the protein-lipid interface ($\theta$) with the current radius of the channel, that is, there is some unknown function $\theta(R)$ \citep{Rees2002}.  Future experiments will be necessary to further clarify this point.  If we make the simplest approximation that $\theta(R)=\mbox{constant}$ and look at two reasonable values of $\theta=0.6$ and $\theta=0.8$ \citep{Turner2004}, using eqn.~\ref{midnucleation} and the parameters in Table \ref{tb1}, we find the rather small critical tensions $\sim 0.004\,k_BT/\mbox{nm}^2$ and $\sim 0.06\,k_BT/\mbox{nm}^2$, respectively, compared to the known critical tension of MscL at $\sim2.5\,k_BT/\mbox{nm}^2$ \citep{SukharevBJ2004, Anishkin2005}.  Though we have shown that midplane deformations are capable of endowing a channel protein with bistability, the scale of the critical tension and the free energy difference between conformations indicates that, at least for MscL, an additional kind of deformation might be important as well.
\n
\begin{center}
\begin{table}
\begin{center}
\begin{tabular}{|l|l|l|}
\hline
Fixed Parameters: & Dynamic Parameter: & Free Energy Difference:\\
\hline
$R=3$ nm, $\theta=0.5$	&	$\tau=0\rightarrow2\,k_BT/\mbox{nm}^2$	&	$10\,k_BT$	\\
\hline
$R=3$ nm, $\tau=2\,k_BT/\mbox{nm}^2$	&	$\theta=0\rightarrow0.5$	&	$10\,k_BT$\\
\hline
$R=3$ nm, $\tau=2\,k_BT/\mbox{nm}^2$	&	$\theta=0\rightarrow0.8$	&	$26\,k_BT$\\
\hline
$\theta=0.5$, $\tau=2\,k_BT/\mbox{nm}^2$	&	$R=3\rightarrow6$ nm	&	$14\,k_BT$\\
\hline
$\theta=0.8$, $\tau=2.5\,k_BT/\mbox{nm}^2$	&	$R=2.5\rightarrow3.5$ nm	&	$13\,k_BT$\\
\hline
\end{tabular}
\caption{Typical free energies for midplane deformation.  The first row indicates how tension leads to an increase in deformation energy.  The second and third rows show the sensitivity to the boundary slope. The fourth row indicates how protein radius changes deformation energy.  The last row is a comparison with the known radius change and critical tension of MscL. }
\label{tb2}
\end{center}
\end{table}
\end{center}

\subsubsection{Thickness Deformations}
\label{leafdef}

We have examined how protein conformation can alter midplane bending of the surrounding lipid bilayer and how this deformation energy penalizes the open state by virtue of its larger radius.  A second major class of deformations  are those that bend and compress a single leaflet of the membrane \citep{Huang1986,Dan1996,AndersenBJ1998,Wiggins2005} and can be thought of as imposing a local thickness on the lipid bilayer that is different from its equilibrium value, as illustrated in Figure \ref{defmodes}.  This kind of deformation relies on the fact that most proteins are rigid in comparison to the flexibility of a lipid molecule.  Hence, when trying to match the hydrophobic region of the protein to the hydrophobic core of the bilayer, it is the lipid that will undergo the vast majority of the deformation.    For the calculations considered here, we assume that leaflet deformations are symmetric:  whatever happens to the top leaflet is mirrored in the bottom leaflet.  The deformation is measured as the deviation of the equilibrium position of the lipid head-groups by the function $u({\bf r})$ at each position ${\bf r}$ on the membrane as was introduced
schematically in  Figure \ref{defmodes}.  The bending energy takes the form 
\be{
G^{(\mbox{\tiny leaf})}_{\mbox{\tiny bend}}=\frac{\kappa_b}{4}\int(\nabla^2 u({\bf r})-c_o)^2\dd^2{\bf r},
}\ee
where $\kappa_b\simeq20\,k_BT$ is the bending modulus of a bilayer \citep{Helfrich1995,Evans2000}, equal to approximately twice the bending modulus of a leaflet, and the spontaneous curvature of the leaflet, $c_o$, characterizes the leaflet's natural tendency for a curved state at a hydrophobic-hydrophilic interface \citep{Dan1998}.  For many bilayer forming lipids, such as phosphatidylcholines, the spontaneous curvature is small \citep{Boal}.  In addition to bending, matching the hydrophobic regions of the protein and bilayer necessarily means the bilayer will change in thickness, giving rise to a bilayer energy penalty of the form
\be{
G^{(\mbox{\tiny leaf})}_{\mbox{\tiny comp}}=\frac{K_A}{2}\int\left(\frac{u({\bf r})}{l}\right)^2\dd^2{\bf r},
}\ee
where  $l\simeq1.75\,\mbox{nm}$ is the leaflet hydrophobic thickness, and due to membrane volume conservation, the bilayer area stretch modulus,  $K_A\simeq60\,k_BT/\mbox{nm}^2$, is associated with this deformation \citep{Evans2000}.   Yet another contribution to the free energy of deformation in those cases where the membrane thickness is perturbed is a local change in the area per lipid as the bilayer thickness varies around the protein. Membrane volume conservation arises because the  membrane is roughly forty times more resistant to volume change than area change \citep{Tosh1986,Seemann2003}.  As a result,  if a transmembrane protein locally thins the bilayer, lipids will suffer an area expansion in a way that conserves volume.  Similarly, if the protein locally thickens the bilayer, the area per lipid will locally decrease.  This implies that  the area change near the protein is proportional to the compression $u(\bf{r})$, and the work done on the bilayer is the integrated area change multiplied by tension
\be{
G^{(\mbox{\tiny leaf})}_{\mbox{\tiny ten}}=\tau\int\frac{u({\bf r})}{l}\dd^2{\bf r},
}\ee
where $\tau$ is the externally applied bilayer tension.  Hence, $u$ less than zero corresponds to a reduction in the energy of the loading device.  All of these contributions can be added up to yield the free energy cost associated with thickness variations of the two leaflets that can be written as
\be{
G^{(\mbox{\tiny leaf})}=\int_A\left({K_A \over 2}\left(\frac{u}{l}\right)^2+\frac{\tau u}{l}+{\kappa_b \over 2}
\left(\nabla^2u\right)^2\right)\dd^2{\bf r}.
\label{gleaf}
}\ee
In elastic models of this type, the protein  dictates the degree of hydrophobic height mismatch, $u(R)=u_o$, and the angle at which the leaflet contacts the protein at the interface between the protein and the surrounding lipids.  Far from the protein, we expect the bilayer to be flat and slightly thinner in accordance with the applied tension, {\it i.e.} $|\nabla u(\infty)|=0$ and $u(\infty)=-\tau l / K_A$, respectively.  In the case of a cylindrical protein we make the further simplifying assumption that the angle is  zero (\ie $|\nabla u(R)|=0$) \citep{Huang1986}. The hydrophobic mismatch itself depends on membrane properties; changes in membrane thickness are linearly related to the hydrophobic mismatch by $u_o=d/2-l$, where $d$ is the hydrophobic thickness of the protein. Unlike midplane deformation, the length scale at which the leaflet returns to its unperturbed state, $\lambda$, depends only on fixed elastic parameters of the membrane given by
\be{
\lambda=\left(\frac{l^2\kappa_b}{K_A}\right)^{\frac{1}{4}}\simeq1\,\mbox{nm}.
}\ee
The deformation energy due to thickness variation in the surrounding lipids induced by the protein can be written in a simple form when the radius of the protein is larger than $\lambda$ (which is the case for MscL) as 
\be{
G^{(\mbox{\tiny leaf})}(R,\tau)=\pi\kappa_b\left(\frac{u_o}{\lambda}+\frac{\tau}{K_A}\frac{l}{\lambda}\right)^2\left(1+\sqrt{2}\frac{R}{\lambda}\right).
\label{Gleafenergy}
}\ee
The deformation energy scales linearly with protein radius and depends quadratically on the hydrophobic mismatch, $u_o$ \citep{Wiggins2005}, making the overall deformation energy particularly sensitive to the hydrophobic mismatch, and hence leaflet thickness $l$ \footnote[3]{The concept of hydrophobic mismatch is valid when the hydrophobic regions of the protein and the bilayer strongly interact, however, this concept has its limits based on the chemistry between the lipids and the transmembrane region of the protein \citep{Sachs2004,Lee2003,Lee2005}, and eventually this condition will be broken if the mismatch is too large \citep{Wiggins2005}.}.  The deformation energy is fairly insensitive to changes in stretch stiffness, $K_A$ ({\it i.e.} most terms in the energy are sublinear), and generally insensitive to changes in the bending modulus since $G\propto\kappa_b^{1/4}$.  Additionally, given the actual values of the elastic parameters, one finds that the leaflet free energy scales roughly {\it linearly} with tension, due to the very small value of $\tau/K_A$.  Like midplane deformation, we see that thickness deformation prefers a smaller protein radius.  On the other hand,  in the midplane case,  tension always increases the deformation energy around a channel while  in the case of lipid bilayer thickness variations,  the tension can either increase or decrease the deformation energy depending on the sign of the hydrophobic mismatch.  In fact, since the hydrophobic mismatch can be either positive or negative (\ie  the protein can be thicker or thinner than the bilayer), tension will increase the deformation energy around a protein that is thicker than the membrane ({\it e.g.~}the closed state of MscL) and decrease the deformation energy around a protein that is thinner than the membrane ({\it e.g.~}the open state), as was shown in Figure \ref{defmodes}.

\begin{figure}
\centering
%\setbmp{-9cm}{0cm}{7cm}{bistable.jpg}
\includegraphics[width=8cm]{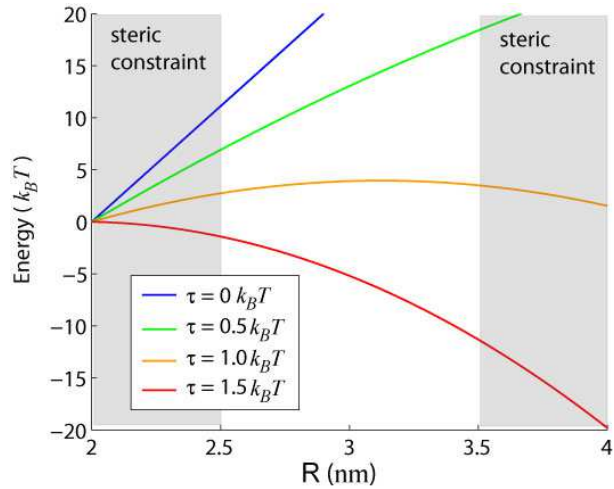}
\caption{Thickness deformation and tension induced energy of a MscL-like channel.  Competition between the cost of deforming the lipid surrounding a protein and the benefit of opening a pore under tension leads to a bistable switch.  At zero tension, the cost of deformation favors a small protein radius, limited only by the steric constraints of the protein.  As tension increases, the benefit to opening a pore is comparable to the energetic cost to deform the lipids surrounding the protein, and a larger protein radius is now possible.  At high tension, the potential energy of the loading device far outweighs the deformation cost and a larger protein radius is favored, again limited by the steric constraints of the protein. }
\label{nucleation}
\end{figure}

One of the beautiful outcomes of this simple thickness variation elastic theory is that the total free energy as a function of protein radius can be written in the simple form
\be{
G(R,\tau)=\underbrace{G^{(\mbox{\tiny leaf})}(R,\tau)}_{\mbox{\footnotesize membrane}}-\underbrace{\tau\pi R^2}_{\mbox{\footnotesize loading device}},
}\ee
which is reminiscent of classical nucleation theory and results in free energy profiles as shown in Figure \ref{nucleation}. At zero tension, the deformation clearly prefers a smaller protein radius, limited only by the steric constraints of the protein structure, which means that there is a certain minimum radius that the protein can adopt.  As the tension increases, the quadratic dependence of the driving force on radius will eventually overcome the linear dependence of the deformation energy, leading to a preference for the open state (corresponding to larger $R$).  We introduce a ``hard wall'' potential at the open radius which provides a severe energy penalty for radii larger than the open state radius and argue that this approximation captures the idea that opening the channel any further would lead to some degree of energetically costly denaturation.  This model also captures the correct scale for the critical gating tension which is on the order of $1\,k_BT/\mbox{nm}^2$.

It is of interest to compare the energy scale implied by this elastic model to measured values.  The free energy change of MscL gating was measured to be $\simeq51\,k_BT$ using native bacterial membranes \citep{SukharevBJ2004, Anishkin2005}.  If one uses the independently measured geometrical properties of the channel, contained in Table \ref{tb1}, and elastic properties of pure bilayers (in the text) to calculate the free energy of the closed and open states, their difference is approximately $55\,k_BT$ at the critical tension of $2.5\,k_BT/\mbox{nm}^2$.    Though very encouraging, this close correspondence depends sensitively upon the choice of hydrophobic mismatch, as dictated by the channel structure and bilayer thickness.
%It is worth noting that, like the midplane bending case, compression deformation also is accompanied by a constraint which relates the hydrophobic mismatch of the channel to the radius ({\it i.e.} $u_o(R)$).  In calculating the free energy change of MscL above, we only made use of the end points of this function at $u_o(R_{closed})$ and $u_o(R_{open})$.

\subsection{Approximating Bilayer Deformation:  The Variational Approach}

In previous sections, we performed cursory derivations of the energy functionals which govern membrane shape for both midplane and membrane thickness deformations.  In order to extract meaning from these energies, we had to minimize the free energy functionals of eqns.~\ref{gmid} and \ref{gleaf} with respect to membrane shape.   To solve the full problem, the conventional scheme (used to obtain the earlier quoted results) is to use the calculus of variations to derive a corresponding partial differential equation in the unknown deformation fields $h({\bf r})$ and $u({\bf r})$.  A useful and intuitive alternative is to adopt a variational approach in which we guess a family of solutions (called `trial functions') that depend upon a small set of parameters and then
minimize the deformation energy with respect to those parameters.

For simplicity, we will showcase this method for one-dimensional membranes which amounts to the approximation that  the protein radius is larger than the natural length-scale of deformation, schematized in Figure \ref{Membrane1D}.  We will use the variational approach to find an approximation for the functions $h({\bf r})$ and $u({\bf r})$ with their related energies, and in the process derive the natural length-scale of deformations in both cases.    Picking a `good' trial function is intimately related to the success of the variational approach.  The choice of the trial function is often dictated by what we know about the character of the solution.  In this case, we know that  in  the near-field the protein is locally disturbing the bilayer by inducing bending or hydrophobic mismatch.  In the far-field, these disturbances should decay back to a flat bilayer.  Keeping in mind that most of the energy cost is stored in the local disturbance around the protein, we want a trial function that has locally varying character around the protein and then a simple decay far from the protein.  Such a trial function (call it $f(x)$) could be constructed using a local disturbance, $g(x)$, within a decaying envelope
\be{
f(x)=g(x)e^{-x/\lambda}.
}\ee
The constant $\lambda$ is an as-yet undetermined natural length scale of deformation and will emerge from the minimization process itself.  Further, this  choice of an exponential envelope essentially guarantees that the  membrane returns to its unperturbed state far from the protein.

As a practical tool for calculation, our choice of $g(x)$ should have enough parameters to reproduce the given boundary conditions.  In addition, we want to choose  $g(x)$ such that the free energy is a simple function of these parameters.  The power of the variational approach is that once we have written the energy in terms of these variational parameters,  the best version of $f(x)$ is, by definition, the one that minimizes the energy.  Thus, for instance, if the trial function has two free parameters $a$ and $b$, $f(x;a,b)$, finding the best trial function amounts to solving a system of algebraic equations defined by
\be{
\frac{\partial}{\partial a} G[f(x;a,b)]=0 \,\,\,\,\mbox{and}\,\,\,\, \frac{\partial}{\partial b}G[f(x;a,b)]=0,
}\ee

\n
where the brackets indicate the energy, $G$, is calculated using the trial function $f$.  This variational strategy can also be used as the basis of numerical approaches in which the membrane
deformation is represented using finite elements, for example.  In this case, the trial functions permit us to determine the energy to an arbitrary degree of accuracy.  Our strategy in the remainder of this section is to use the simplicity of the variational approach to find approximate energies for the midplane and thickness deformations imposed by membrane proteins.

\subsubsection{Variational Approach for Midplane Deformations}

\begin{figure}
\centering
%\setbmp{-13cm}{0cm}{2.5cm}{membranelinetension.jpg}
\includegraphics[width=13cm]{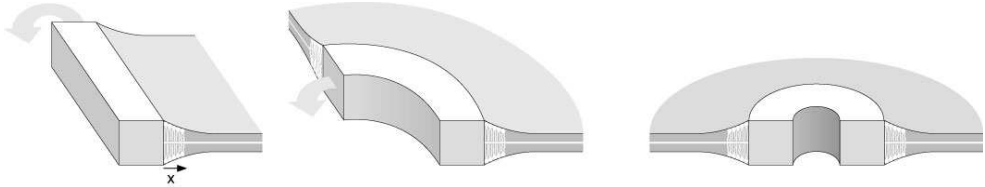}
\caption{Protein-induced line tension.  Deformation of the membrane around an ion channel can be described using a line tension.  This line tension is obtained by solving for a one-dimensional deformation energy per unit length and then imposing that energy around the circumference of the channel.  The diagrams above show the sequential wrapping of this one-dimensional line tension around a cylindrical channel.  }
\label{Membrane1D}
\end{figure}

Our goal is to obtain an approximate expression for the one-dimensional energy due to midplane bending given by 
\be{
G^{(\mbox{\tiny mid})}=2 \pi R\int_0^{\infty}\left({\tau \over 2}\left(\frac{\dd}{\dd x} h(x)\right)^2+
{\kappa_b \over 2} \left(\frac{\dd^2}{\dd x^2}h(x)\right)^2\right)\dd x.
\label{gmidvar}
}\ee
The presence of the $2 \pi R$ in this expression is due to the fact that we are computing the energy per unit length for a deformed bilayer, as shown in Figure \ref{Membrane1D}, and must then multiply by the length (the circumference) of deformed material.

The strategy employed in the variational approach is to plug the trial function into the free energy functional and compute the resulting energy, which depends upon the parameters in the trial function.  Our trial function has the form
\be{
h(x)=g(x)e^{-\left(\frac{x}{\lambda}\right)}.
}\ee
The choice of $g(x)$ can be made based upon the boundary conditions. In particular,  at the boundary of the protein, we require that 
\be{
\frac{\dd}{\dd x}h(x)|_{x=0}=\theta,
}\ee
which tells us that we can make the choice $g(x)=\mbox{constant}$.  Applying this boundary condition yields the functional form
\be{
h(x)=-\theta\lambda e^{-\left(\frac{x}{\lambda}\right)},
}\ee
where the only remaining undetermined  parameter  is the  length scale, $\lambda$.  This trial function can be plugged into eqn.~\ref{gmidvar} and the integral is easily evaluated to yield the free energy
\be{
G^{(\mbox{\tiny mid})}(\lambda)=\frac{\pi}{2}R\theta^2\left(\tau\lambda+\frac{\kappa_b}{\lambda}\right)
}.
\ee
The next step in the variational strategy is to minimize the free energy with respect to $\lambda$,
\be{
\frac{\partial}{\partial \lambda} G^{(\mbox{\tiny mid})}(\lambda)=0 \,\,\,\, \rightarrow\,\,\,\, \lambda=\sqrt{\frac{\kappa_b}{\tau}},
}\ee
which upon substitution yields
\be{
G^{(\mbox{\tiny mid})}=\theta^2\pi\kappa_b\frac{R}{\lambda}=\theta^2\pi R\sqrt{\kappa_b\tau}.
}\ee
This is precisely the asymptotic ($R\sqrt{\tau/\kappa_b}>1$) form of eqn.~\ref{gmidenergy} for midplane bending energy, and our minimization correctly defines the natural length-scale of midplane deformation.

\subsubsection{Variational Approach for Membrane Thickness Deformations}

A similar analysis can be made for the one-dimensional deformations induced by hydrophobic mismatch. In this case, the free energy functional in the absence of tension  can be written as 
\be{
G^{(\mbox{\tiny leaf})}=2\pi R\int_0^{\infty}\left({K_A \over 2}\left(\frac{u(x)}{l}\right)^2+
{\kappa_b \over 2}\left(\frac{\dd^2}{\dd x^2}u(x)\right)^2\right)\dd x.
\label{gleafvar}
}\ee
We adopt the same functional form for the trial function, namely, 
\be{
u(x)=g(x)e^{-\left(\frac{x}{\lambda}\right)}.
}\ee

\n
In this case, we specify two boundary conditions in the near-field; there is a hydrophobic mismatch which demands
\be{
u(R)=u_o,
}\ee

\n
and the leaflet has a particular slope at the membrane interface, which we will set to zero,
\be{
\frac{\dd}{\dd x}u(x)|_{x=0}=0.
}\ee

\n
In order to accommodate these two boundary conditions, $g(x)$  must have two free parameters.  As a result,  we pick the simplest function which has two degrees of freedom, namely a line, and hence set $g(x)=ax/\lambda+b$, where $a$ and $b$ are constants.  Applying the two near-field boundary conditions constrains the trial function to the form
\be{
u(x)=u_o\left(1+\frac{x}{\lambda}\right)e^{-\left(\frac{x}{\lambda}\right)},
}\ee

\n
where $\lambda$ is a free parameter with respect to which the energy must be minimized.  Using this trial function, the free energy can be written as a simple expression of the form
\be{
G^{(\mbox{\tiny leaf})}(\lambda)=\pi\kappa_b u_o^2 R\left(\frac{5}{4}\frac{K_A}{\kappa_bl^2}\lambda+\frac{1}{4\lambda^3}\right).
}\ee

\n
Minimizing the energy with respect to $\lambda$ gives
\be{
\frac{\partial}{\partial \lambda} G^{(\mbox{\tiny leaf})}(\lambda)=0 \,\,\,\, \rightarrow\,\,\,\, \lambda=\left(\frac{3}{5}\right)^{\frac{1}{4}}\left(\frac{\kappa_bl^2}{K_A}\right)^{\frac{1}{4}}
}\ee

\n
which upon substitution gives the membrane thickness  energy
\be{
G^{(\mbox{\tiny leaf})}=\left(\frac{5}{3}\right)^{\frac{3}{4}}\left(\frac{K_A}{\kappa_b l^2}\right)^{\frac{3}{4}}\pi\kappa_b u_o^2 R.
}\ee

\n
Again, the variational approach has reproduced the correct asymptotic form of the energy with a small multiplicative error (see eqn.~\ref{Gleafenergy}); the exact asymptotic result  has $\sqrt{2}$ instead of $\left(\frac{5}{3}\right)^{\frac{3}{4}}$, introducing an error of $\sim4\%$.  

\begin{figure}
\centering
%\setbmp{-13cm}{0cm}{3cm}{trialfunc.jpg}
\includegraphics[width=13cm]{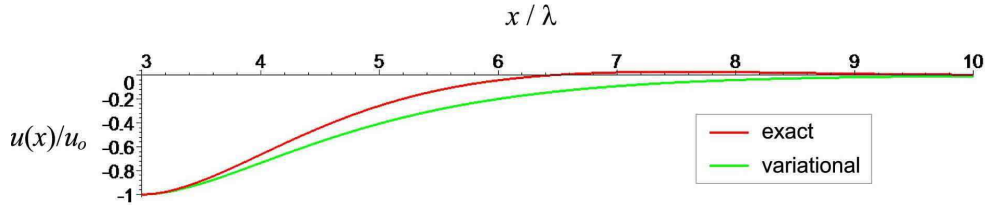}
\caption{Comparison of the exact and variational solutions for the thickness deformations around a protein. The variational approach generates an approximation to $u(x)$ which is close to the exact solution.  The protein radius is $R/\lambda=3$. }
\label{trial}
\end{figure}

Finally, there are many forms of $u(x)$ which yield roughly the same {\it energy}, but how does the exact deformation {\it shape} compare with our minimized trial function?  Here too, the variational approach gives a trial function that nearly matches the exact result as shown in Figure \ref{trial}.

\subsection{Distilling the Design Principles}

\begin{figure}
\begin{center}
%\setbmp{-9cm}{0cm}{6cm}{mscLdeformationenergies.jpg}
\includegraphics[width=10cm]{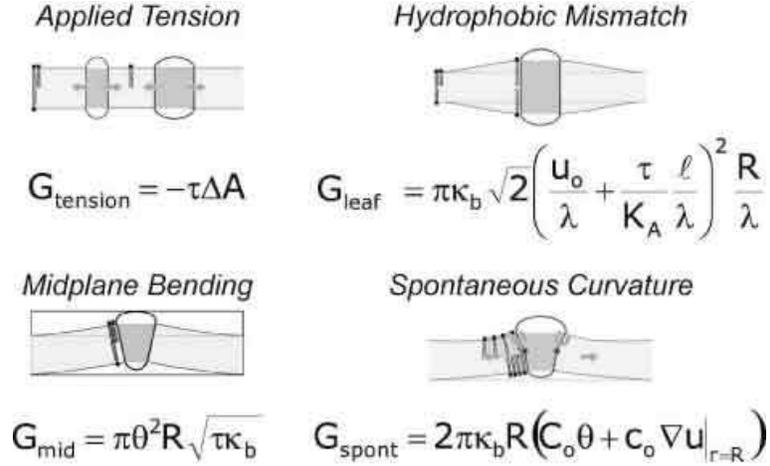}
\caption{Contributions to the free energy.  This figure shows how the different modes of deformation contribute to the overall free energy budget of the membrane-protein system.  The energies are written asymptotically to show their dominant scaling with the relevant parameters. For the sake of simplicity, we did not address how spontaneous curvature factors into the free energy budget.  However, a thorough discussion of both midplane and leaflet spontaneous curvature energy contributions are found in \protect\citet{Wiggins2005}.}
\label{MscLDeconstruct}
\end{center}
\end{figure}

Having explored how midplane bending, thickness variation  and area change are coupled to tension and the geometric features of the MscL channel, can we distill general rules for what makes a membrane protein mechanosensitive?  One simple statement is that under tension an increase in protein area is always favored,  regardless of bilayer elastic properties, because an increase in area lowers the potential energy of the loading device.  Conversely, both midplane and thickness deformations prefer a smaller channel, because a larger radius results in a larger annulus of deformed lipid and hence a larger free energy penalty (except in the case where the spontaneous curvature favors a larger radius \citep{Wiggins2005}). With the area change preferring a larger radius and deformation preferring a smaller radius, we have the necessary energetic competition that ultimately leads to bistability.  This also means the sign of the free energy change due to deformation (midplane or thickness) must be positive.  Hence, the channel is going from a closed state with less deformed lipid surrounding it, to an open state with more deformed lipid surrounding it.  The contributions to the free energy budget of a mechanosensitive protein, like MscL, due to channel-area change and membrane deformations are shown in Figure \ref{MscLDeconstruct}.  The basic point of this picture is to show how various contributions to the free energy scale with the radius ($R$) and the elastic parameters.

Midplane deformation is the deformation which depends most simply on membrane properties, since it is only linked to the bending modulus.  Additionally, its tension dependence is such that the cost of the deformation always increases with tension and angle, hence we know that tension in addition to preferring a larger protein, also wants a more cylindrical protein in the case of midplane bending.  This allows us, within the limitations of our theory, to put an upper bound on the cost of midplane deformations.  Taking the lytic tension as an upper bound, a nominal bending modulus of $20\,k_BT$, and $\theta=0.6$ as a reasonable value of the membrane slope \citep{Turner2004}, the maximum energetic cost of deformations for a protein of radius $R$ (in nm) is $\simeq R\times9\,k_BT/\mbox{nm}$.

Thickness  deformation depends on all the elastic parameters; bending modulus, area stretch modulus, and membrane thickness.  The tension dependence of thickness deformation energy is also more complex, though a general principle does emerge.  We know that tension can increase or decrease the overall thickness deformation energy, but the general principle is that it always prefers the protein to have the same hydrophobic thickness as the bilayer, though the bilayer thickness is itself decreased as tension increases.  The other important feature to note is that a decrease in the thickness of a transmembrane protein is always accompanied by an increase in the area of the membrane surrounding the protein due to volume conservation of the membrane.  This change in membrane area is {\it indistinguishable} from a change in protein area.  Indeed for MscL, the measured area change is probably a mix of a change in the areal footprint of the protein, and a local increase in the membrane area surrounding the protein, together giving the measured value of $\sim20\,\mbox{nm}^2$. An estimate of the upper bound of leaflet deformations is made by assuming the maximum $u_o=0.5\,\mbox{nm}$, then the maximum change in free energy for a protein of radius $R$ (in nm) is $\simeq R\times22\,k_BT/\mbox{nm}$ at zero tension with the given elastic parameters (see Table \ref{tb1}).  This illustrates that while both midplane and thickness  deformations are important factors in determining the preferred protein conformation, thickness deformations are generally  associated with a slightly higher energy scale.

\section{Experimental Considerations}

\begin{figure}
\centering
%\setbmp{-14cm}{0cm}{7.5cm}{pipette.jpg}
\includegraphics[width=14.5cm]{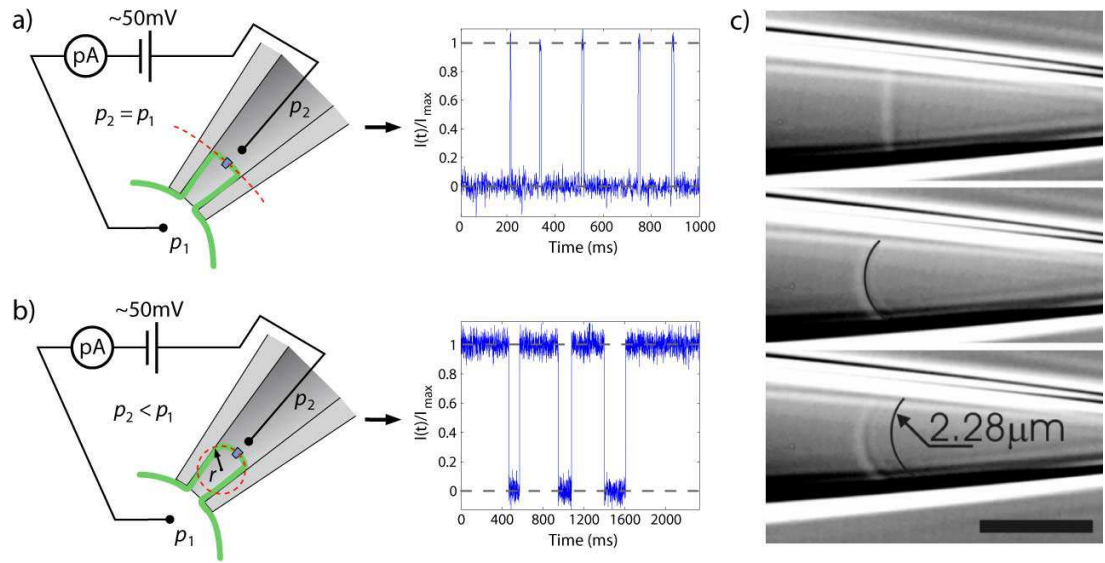}
\caption{Measurement of tension vs. pressure difference in an electrophysiological experiment.   A channel protein (small blue rectangle) is embedded in a membrane patch (green). A potential of roughly 50mV is applied across the sealed membrane patch, and channel opening events are measured by an ammeter (circle) with picoamp (pA) sensitivity. a) At low pressure difference, the tension in the patch is low, the mechanosensitive channel is in the closed conformation, and the patch has a very large radius of curvature.  The plot to the right shows normalized channel current as a function of time for a simulated channel; the open state has low occupation at low tension. b) At high pressure difference, the tension in the patch is high, the mechanosensitive channel will occupy the open state, and the radius of curvature (r) is on the order of microns.  The plot to the right shows the open state has high occupation at high tension. c)  Optical micrograph of vertically oriented membrane patch at low (top) and high (middle and bottom) pressure differences, illustrating the decrease in the radius of curvature with increase pressure difference (from \protect\citet{Moeblount2005}).  The scale bar is $5\mu\mbox{m}$.}
\label{pipette}
\end{figure}

Much of our knowledge of the function of mechanosensitive channels, including MscL, comes from detailed electrophysiology studies where gating of the channel is monitored by sharp differences in the ion flux through a membrane patch \citep{SukharevAnnRev1997,SukharevJGP1999,SukharevNature2001,PerozoNSB2002,SukharevBJ2004, Anishkin2005}.  A small voltage ($\sim50$mV) is applied across a patch of membrane at the tip of a micropipette.  As a function of pressure difference, channel opening events are recorded as stochastic changes in patch current by an ammeter with picoamp (pA) sensitivity.  This truly amazing single-molecule  spectroscopy technique allows the experimenter to adjust the voltage  as well as the pressure difference across the membrane as shown in  Figure \ref{pipette}.  The pressure difference across the membrane translates into a lateral membrane tension (via the Laplace-Young Relation), responsible for gating the mechanosensitive channel. However, there are two serious problems with this method when probing the mechanisms of mechanosensitive channels. 

Arguably, the most serious problem is that often {\it pressure difference} (J/m$^3$) across the membrane is taken to be the input variable of prime importance, when in fact {\it tension} (J/m$^2$) is the membrane parameter which governs mechanosensitive gating.  Pressure difference is linearly related to tension via the radius of curvature of the membrane, hence in principle the fix is straightforward - image the membrane patch (see Figure \ref{pipette}).  While certainly not impossible \citep{SukharevJGP1999,Moeblount2005}, the membrane patch can be difficult to image due to its small size and the fact that it is inside the micropipette.  A recent study \citep{Moeblount2005} demonstrated the importance of measuring tension in lieu of pressure difference.  It was shown that using the standard methods for creating ``identical'' micropipettes, the measured characteristics of a channel varied significantly.  However, when the membrane patch was imaged and tension used as the principle input variable, the same data collapsed to within a few percent of each other.  In general, if one could perfectly control the size and shape of the micropipette tip used for contacting and sealing the membrane patch, all measurements would be related by a single constant (the radius of curvature).  However, variations in micropipette shape and size, as well as variations in how the membrane contacts the pipette tip all lead to potentially large variations in the perceived gating characteristics of the channel.  Additionally, it is difficult to compare the wealth of quantitative data coming from electrophysiology studies to theoretical models when pressure difference, instead of tension, is used as the principle input variable.  Tension is routinely measured in micropipette aspiration experiments \citep{Evans2000}, and in fact, single-channel electrophysiology recordings are possible in such a setup \citep{GoulianBJ1998} using ion channels with conductances {\it lower} than MscL.  Hence, this technique might provide a useful way to apply known membrane tension to reconstituted MscL channels in well characterized membranes.

With tension being used as the variable of prime importance, electrophysiology is poised to put the continuum mechanical view to the test, elucidating the role of lipids in ion channel function.  In particular, the elastic properties of many lipids have been measured \citep{Evans2000}, enabling a careful examination of the dependence of gating energy on lipid carbon chain length.  The simple continuum view we set forth here predicts a quadratic dependence of the lipid thickness deformation energy on hydrophobic mismatch, which is directly linked to carbon chain length.  This, of course, has implications for both the function of various transmembrane proteins, and comments meaningfully on the ability of bilayer thickness to segregate proteins in biological membranes.

A second class of intriguing experiments concerns the mechanosensitivity of other ion channels and receptors, generally regarded not to be mechanosensitive \citep{MorrisBJ2001,MorrisBJ2002}.  This is both interesting from a functional standpoint, in an effort to understand the full physiological effects of these proteins, and as a tool for understanding  structural features such as the motions of transmembrane helices.  Performing a similar experiment where lipid carbon chain length is varied around a voltage-gated ion channel (for example) could reveal hidden mechanosensitivity, and energetic analysis from such an experiment could comment on the degree of height and area change during the gating transition.

The second problem facing a complete understanding of the function of mechanosensitive channels is that for many such channels volumetric flow, and not ion flux, is the relevant physiological parameter\footnote[2]{The mechanosensitive bacterial channel MscS \citep{Pivetti2003,Levina1999,Bass2002} is another example.  Although, there are also mechanosensitive channels that appear to be highly ion selective, such as the bacterial mechanosensitive ion channel MscK \citep{Blountembo2002} and the K2P family of mammalian mechanosensitive channels \citep{Maingret1999,Maingret2000,Franks2004,Lauritzen2005}.}. Hence, ion flux is used as a surrogate measurable in place of the true physiological output of the channel.  One could argue that ion flux is proportional to volumetric flow, however this assumes that the way ions flow through the channel pore is identical to the way water flows through the pore.  Experiments have elucidated the roughly ohmic nature of mechanosensitive channels \citep{PerozoNSB2002, Martinac1997} at low voltage ($\lesssim80\mbox{mV}$), however we know essentially nothing about how a pressure gradient across the membrane translates into a volumetric flow.  Even the simplest continuum approximation (Hagen-Poiseuille flow) would predict a non-linear function relating the area of the channel pore to the volumetric flow, in contrast to the (roughly) linear relationship between ion flux and channel pore area as predicted by Ohm's Law \citep{Hille1968}. It would be of considerable physical and physiological interest to expand our understanding of fluid flow at the molecular level, by measuring the relationship between pressure gradient and volumetric flow through a large-pore channel like MscL.

\section{Cooperativity and Interaction between Transmembrane Proteins}

One intriguing consequence of the deformations induced in membranes by ion channels is that channels will interact.   These interactions can lead to cooperativity in the gating of  neighboring channels and can also induce spatial ordering of the proteins. These interactions can be thought of as arising from two different effects:   those of {\it elastic} origin and those of {\it thermal} origin.  The elastic forces are purely an enthalpic effect coming from a minimization of the deformation energy around two proteins separated by a given distance.  The thermal forces are entropically driven by the thermal fluctuations of the membrane and are analogous to Van der Waals forces.

\subsection{Enthalpic Interactions}

As discussed above, proteins which change the membrane thickness or bend the membrane midplane produce deformations which extend anywhere from a few nanometers (thickness) up to tens of nanometers (midplane) from the protein edge.  As two proteins approach each other, their respective deformation fields overlap resulting in a deformation profile between them that is different
than either of them produce separately. In this case, the total deformation energy of the system is dependent on the separation between the two proteins and results in an interaction potential which is dependent upon the conformation of the proteins.  These  forces arise purely from the mechanical attributes of the deformed membrane and have no entropic component.  We know that midplane and thickness  deformations are independent, and hence there are distinct interactions due to midplane and thickness deformations.   
 
Pairwise interactions due to midplane deformation using eqn.~\ref{gmid} have previously been  calculated for a variety of membrane curvature environments and protein shapes at zero tension \citep{OsterBJ2001}.  Using a bilayer bending modulus of $\sim100\,k_BT$, attractive interactions of order $\sim 1-5\,k_BT$ were found when the proteins were separated by 1 - 2 protein radii (which we estimate to be 5 - 10 nm measured center-to-center for a typical transmembrane protein).  If we adjust the energy scale to be consistent with a phosphatidylcholine  bilayer bending modulus of $\sim20\,k_BT$ this lowers the interaction energetics to $\sim0.5-3\,k_BT$.  These interactions tend to be long-ranged with a power-law decay of $1/r^4$ \citep{Goulian1993}.    Simple pairwise interaction will be inadequate to describe the nature of interactions between more than two proteins.   This arises because one protein can shield other proteins from feeling the deformation of a neighboring protein, and hence interactions are not (in general) pairwise additive. Apart from direct numerical simulation, there are few analytical (theoretical) tools which allow one to study how many interacting proteins in close proximity behave as a group \citep{HuangBJ1999}.  
  
\begin{figure}
\centering
%\setbmp{-13.5cm}{0cm}{6cm}{twochannelconfig.jpg}
\includegraphics[width=14cm]{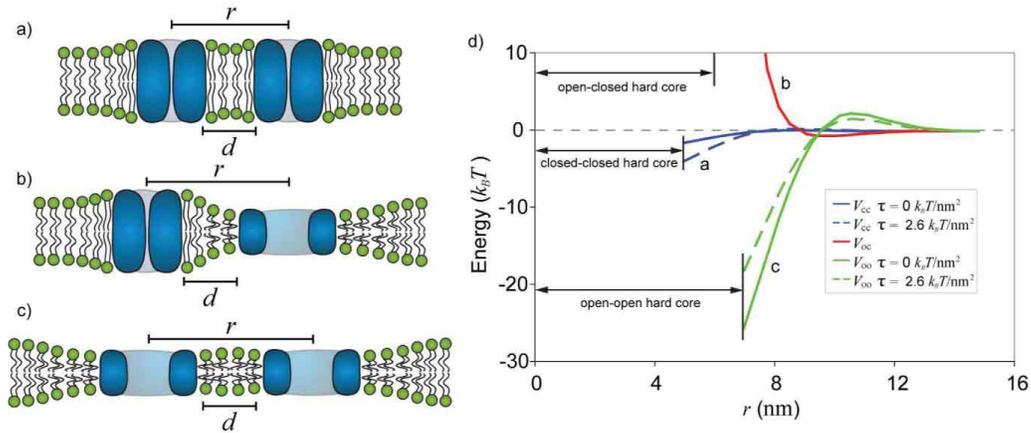}
\caption{Conformation-dependent interactions between two MscL channels.   As two MscL channels (blue) come close to each other, regions of deformed lipids (green) overlap leading to deformation induced interactions. a) Deformation surrounding two closed MscL channels. b) Deformation surrounding a closed and an open MscL channel. c) Deformation surrounding two open MscL channels.  The relative sizes of the open channel, the closed channel, and the lipids are roughly correct.  d)  Interaction potentials for the three configurations shown in a, b and c.  External tension weakens the interaction between two open channels ($V_{\mbox{\tiny oo}}$) and strengthens the interaction between two closed channels ($V_{\mbox{\tiny cc}}$), but has almost no effect on the interaction between an open and closed channel ($V_{\mbox{\tiny oc}}$).  The open-open and closed-closed interactions are both more strongly attractive than the open-closed interaction, indicating that elastic potentials favor interactions between channels in the same state. The `hard core' distance is where the proteins' edges are in contact.}
\label{Msclint}
\end{figure}

Like midplane deformations, the thickness  deformation fields extending from the edges of two proteins will overlap and interact as the proteins come into close proximity \citep{Dan1993, Dan1996}.  We provided evidence that lipids likely influence the function of MscL through thickness deformations and once again we will  appeal to MscL as a case study for interacting membrane proteins.  The short-range nature of thickness deformations (essentially exponential decay) means there is no power-law asymptotic formula for their interaction, though we numerically explored these interactions for all possible conformations of two MscL proteins as shown in Figure \ref{Msclint}.  As we saw with single proteins, the energetic scale of thickness  interactions is generally higher than with midplane deformations, and can vary greatly depending on the hydrophobic mismatch.  The leaflet interactions between two MscL proteins are appreciable when they are within several nanometers of each other, and ranged from $\sim2-25\,k_BT$ depending on the protein conformations and the tension in the membrane. This kind of short-ranged interaction might play an important role in membrane protein function \citep{Goforth2003,Molina2006,Botelho2006}, given the nominal density of transmembrane proteins in biological membranes leads to spacings on the order of 10-100 nm \citep{Engelman2004}.

Additionally, the interactions due to thickness variations can be either attractive or repulsive depending on the shape of the proteins.  The general principle that emerges is that `like' proteins attract and `unlike' proteins repel (in contrast to electrostatics) as shown in Figure \ref{Msclint}.  Proteins whose hydrophobic mismatch has the same sign (\ie both are taller or both are shorter than the membrane) lead to net attractive interactions; proteins with opposite signs of hydrophobic mismatch lead to repulsive interactions.   Later in the article, we will demonstrate that these conformation-dependent interactions can communicate state information between two proteins, leading to cooperative channel gating. 

\subsection{Entropic Interactions}

A second class of forces between membrane proteins arise due to membrane fluctuations.  Like most entropic forces, the thermal interactions between transmembrane proteins are fairly weak, on the order of a few $k_BT$.  Two fluctuation-induced forces have been studied in some detail in the literature; a long-ranged Casimir force due to the surface fluctuations of the membrane \citep{Goulian1993,ParkJFr1996}, and a very short-ranged depletion force due to the excluded volume of lipid molecules between two membrane proteins \citep{Sintes1997}.  

The Casimir force between two membrane proteins arises because the available spectrum of fluctuations of the membrane-midplane depend on the distance between two proteins.  Entropically, the membrane-protein system seeks to maximize the number of available modes of fluctuation and hence an energetic potential exists between two transmembrane proteins in a fluctuating, thermally active membrane.  Through a series of elegant calculations, this force was shown to have a $1/r^4$ asymptotic form, where $r$ is the center-to-center distance between two cylindrical proteins \citep{Goulian1993, ParkJFr1996}.  If we presume that it is approximately correct for small separations ($r\simeq2R$),  this implies an attractive potential with an energy scale of $\sim1\,k_BT$. 

Lateral density fluctuations of lipids in the membrane also lead to entropic forces between proteins.  Using Monte Carlo simulations, these entropic depletion forces (also called `excluded volume forces') between cylindrical proteins were shown to be appreciable only when the proteins' edges were within $\sim1$ lipid molecular diameter \citep{Sintes1997}.  For cylindrical proteins, with diameters on the order of $\sim1-2\,\mbox{nm}$, direct edge contact resulted in a  favorable interaction with an energy scale of $\sim2\,k_BT$.  

\subsection{Protein Conformations Affected by Interaction}

As noted above, the elastic interactions between ion channels such as MscL depend upon protein conformation.  In earlier sections of the paper,  we established that the equilibrium conformations of the channel are entirely determined by the free energy difference between the two states.  As a result, elastic interactions which change the energy of a two-channel system will affect the probability that we measure any one channel in the open state.  In fact, electrophysiology (see Figure \ref{pipette}) is well suited to such measurements where the total amount of time spent in the open state divided by the total measurement time {\it is} the open probability.

\begin{figure}
\centering
%\setbmp{-9cm}{0cm}{6cm}{popens.jpg}
\includegraphics[width=10cm]{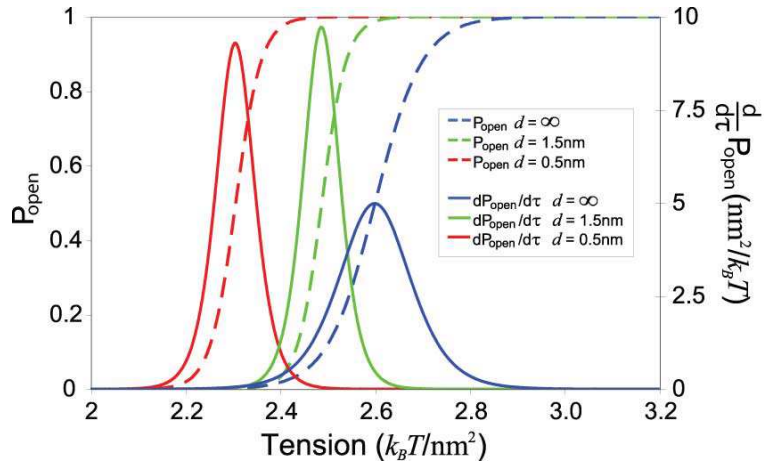}
\caption{Conformational statistics of interacting MscL proteins.  Interactions between neighboring channels lead to shifts in the probability that a channel will be in the open state (dashed lines).  The sensitivity and range of response to tension, ${\mbox{d}P_{\mbox{\tiny open}}/\mbox{d}\tau}$, are also affected by bilayer deformations (solid lines).  ${P_{\mbox{\tiny open}}}$ and ${\mbox{d}P_{\mbox{\tiny open}}/\mbox{d}\tau}$ are shown for separations of $0.5\,\mbox{nm}$ (red) and $1.5\,\mbox{nm}$ (green) with reference to non-interacting channels at $d=\infty$ (blue).  Interactions shift the critical gating tension for the closest separation by $\sim12\%$.  Additionally, the peak sensitivity is increased by $\sim90\%$ from $\sim5\mbox{nm}^2/k_BT$  to $\sim9.5\mbox{nm}^2/k_BT$.}
\label{FigPopen}
\end{figure}

The free energy difference between the open and closed states of a single channel is roughly $50\,k_BT$ \citep{SukharevBJ2004}, which implies the energy scale for two channels is roughly $100\,k_BT$.   We have also seen that two MscL channels in proximity have interactions on an energy scale of roughly $\sim20\,k_BT$ as shown in Figure \ref{Msclint}.   Two open channels have a strong, favorable interaction that can significantly alter the open probability of a given channel relative to the isolated channel value as shown in Figure \ref{FigPopen}.   Such interactions also affect channel `sensitivity', defined by the derivative of the $P_{\mbox{\tiny open}}$ curve with respect to tension, which quantifies how responsive the channel is to changes in the driving force, in this case tension.  The full-width at half maximum of this peaked function is a measure of the range of tension over which the channel has an appreciable response.  In general, the area under the sensitivity curve is equal to one, hence increases in sensitivity are always accompanied by decreases in range of response, as demonstrated by the effects of the beneficial open-open interaction on channel statistics.  The critical gating tension and sensitivity are essentially the key properties which define the transition to the open state, and are analogs to the properties which define the transition of {\it any} two-state ion channel.  Hence, the elastic interactions can affect channel function on a fundamental level.\\

\n
{\bf Conclusion}\\
\n
The goal of this article is to take stock of the role of lipid bilayer deformations in mechanosensation.   More precisely, we have argued that the lipid bilayer is not a passive bystander in the energetics of channel gating.  As a result, by tuning membrane properties it is possible to alter channel function.   We have emphasized two broad classes of membrane deformation that are induced by the presence of a transmembrane protein: i) deformation and bending of the midplane of the lipid bilayer, ii) variations in the thickness of the lipid bilayer that are induced by hydrophobic mismatch.   As a result of these deformations, there is a free energy cost to changing the radius of a channel since the open state implies a larger annulus of deformed material and hence a higher free energy.  This deformation energy competes with the energetic relaxation of the loading device.

One of the key reasons for performing theoretical analyses like those described here is that they permit us to sharpen the questions that can be asked about a given biological problem.    This sharpness is ultimately most meaningful if it is translated into precise experimental predictions.  The theoretical results described here suggest a variety of experimental predictions.

\begin{itemize}
\item{\emph{Dependence of gating tension on hydrophobic mismatch.}   Previous work has already shown that lipid bilayer tail lengths can alter channel gating by changing the hydrophobic mismatch.  To more precisely examine this relationship, careful measurements of the membrane tension need to be made, as opposed to pipette pressures, to elucidate the energetics underlying gating.  Alternatively, mutagenesis could be used to explore the same effect by changing the hydrophobic thickness of the protein.}

\item{\emph{Hidden mechanosensitivity in other classes of channels and receptors.}  The results described here have been applied to the case study of MscL.  However, we argue that any transmembrane protein that varies its radius or hydrophobic thickness upon conformational change will exhibit mechanosensitivity.  Furthermore, the way tension affects the function of these proteins might help elucidate the classes of structural changes that occur during their conformational change.}

\item{\emph{Cooperative gating of channels.}  As a result of the elastic deformations induced in the lipid bilayer by mechanosensitive channels, nearby channels can communicate their conformational state, resulting in cooperative gating.  This cooperativity should be observable in electrophysiology experiments as a change in the critical tension and channel sensitivity with an increase in channel density. }
\end{itemize}

{\it Shortcomings of the Theory.}  Obviously, the use of simple ideas from elasticity theory to capture the complex process of mechanosensation provides a caricature of the real process.   One signature of the shortcomings of this kind of approach is the fact that single amino acid substitutions can completely alter the properties of certain proteins \citep{Yoshimura1999,Yoshimura2004}.  This serves as a warning of the pitfalls of models that ignore atomic-level details and their impact on biological function.   A second class of complaint that can be registered against the models described here is that we have ignored material heterogeneity.  In particular, biological membranes are built up of a broad range of different lipids and are riddled with membrane proteins.  As a result, it is not clear if an elastic description like that  described here is appropriate, and if it is, how to select the relevant material parameters.

Regardless of the difficulties highlighted above, it is clear that the emergence of an increasing number of structures of ion channels coupled with functional studies of these proteins has raised the bar for what should be expected of theoretical models of channel function.   The central thesis of the work described here is that the presence of the lipid bilayer provides another way in which these systems can be manipulated.\\

\n
{\bf Acknowledgments}\\
\n
We are  grateful to a number of people who have given us both guidance and amusement in thinking about these problems, as well as insightful comments on the manuscript:  Doug Rees, Olaf Anderson, Fred Sachs, Evan Evans, Cathy Morris, Sergei Sukharev, Mathew Turner, Eduardo Perozo, Nily Dan, Fyl Pincus, Liz  Haswell and Pierre Sens.  JK acknowledges the support of National Science Foundation grant No. DMR-0403997 and is a Cottrell Scholar of Research Corporation.  RP acknowledges the support of the National Science Foundation grant No. CMS-0301657.  TU and RP acknowledge the support of the National Science Foundation CIMMS Award ACI-0204932 and NIRT Award CMS-0404031 as well as the National Institutes of Health Director's Pioneer Award.  DR acknowledges the support of National Science Foundation grant No. DGE-0549390.  PAW is supported by the Whitehead Institute for Biomedical Research.\\

\n
{\bf Accession Numbers}\\
\n
The primary accession numbers (in parentheses) from the Protein Data Bank (http://www.pdb.org) are:  mechanosensitive channel of large conductance (2OAR; formerly 1MSL).

\bibliographystyle{plainnat}
%\bibliography{springer2007}

\begin{thebibliography}{81}
%There are 81 references.
\expandafter\ifx\csname natexlab\endcsname\relax\def\natexlab#1{#1}\fi
\expandafter\ifx\csname url\endcsname\relax
\def\url#1{{\tt #1}}\fi

\bibitem[Akitake et~al(2005)Akitake, Anishkin, and Sukharev]{Akitake2005}
Akitake B, Anishkin A, Sukharev S (2005)
\newblock The``dashpot" mechanism of stretch-dependent gating in MscS.
\newblock { J Gen Physiol}, 125\penalty0 (2):\penalty0 143--54

\bibitem[Anishkin et~al(2005)Anishkin, Chiang, and Sukharev]{Anishkin2005}
Anishkin A, Chiang CS, Sukharev S (2005)
\newblock Gain-of-function mutations reveal expanded intermediate states and a sequential action of two gates in MscL.
\newblock {J Gen Physiol}, 125\penalty0 (2):\penalty0 155--70

\bibitem[Aranda-Espinoza et~al(1996)Aranda-Espinoza, Berman, Dan, Pincus, and
  Safran]{Dan1996}
Aranda-Espinoza H, Berman A, Dan N, Pincus P, Safran S (1996)
\newblock Interaction between inclusions embedded in membranes.
\newblock {Biophys J}, 71\penalty0 (2):\penalty0 648--56

\bibitem[Barry and Lynch(2005)]{Barry2005}
Barry PH, Lynch JW (2005)
\newblock Ligand-gated channels.
\newblock {IEEE Trans Nanobioscience}, 4\penalty0 (1):\penalty0 70--80

\bibitem[Bass et~al(2002)Bass, Strop, Barclay, and Rees]{Bass2002}
Bass RB, Strop P, Barclay M, Rees DC (2002)
\newblock Crystal structure of {\it Escherichia coli} MscS, a voltage-modulated and  mechanosensitive channel.
\newblock {Science}, 298\penalty0 (5598):\penalty0 1582--7

\bibitem[Boal(2002)]{Boal}
Boal D (2002)
\newblock {Mechanics of the cell}.
\newblock Cambridge University Press, New York, 1st edition

\bibitem[Botelho et~al(2006)Botelho, Huber, Sakmar, and Brown]{Botelho2006}
Botelho AV, Huber T, Sakmar TP, Brown MF (2006)
\newblock Curvature and hydrophobic forces drive oligomerization and modulate activity of Rhodopsin in membranes.
\newblock {Biophys J}, 91\penalty0 (12):\penalty0 4464--77

\bibitem[Calabrese et~al(2002)Calabrese, Tabarean, Juranka, and
  Morris]{MorrisBJ2002}
Calabrese B, Tabarean IV, Juranka P, Morris CE (2002)
\newblock Mechanosensitivity of n-type calcium channel currents.
\newblock {Biophys J}, 83\penalty0 (5):\penalty0 2560--74

\bibitem[Cantor(1999)]{Cantor1999}
Cantor RS (1999)
\newblock Lipid composition and the lateral pressure profile in bilayers.
\newblock {Biophys J}, 76\penalty0 (5):\penalty0 2625--39

\bibitem[Chang et~al(1998)Chang, Spencer, Lee, Barclay, and Rees]{Rees1998}
Chang G, Spencer RH, Lee AT, Barclay MT, Rees DC (1998)
\newblock Structure of the MscL homolog from {\it Mycobacterium tuberculosis}: A gated mechanosensitive ion channel.
\newblock {Science}, 282\penalty0 (5397):\penalty0 2220--6

\bibitem[Chiang et~al(2004)Chiang, Anishkin, and Sukharev]{SukharevBJ2004}
Chiang CS, Anishkin A, Sukharev S (2004)
\newblock Gating of the large mechanosensitive channel in situ: Estimation of the spatial scale of the transition from channel population responses.
\newblock {Biophys J}, 86\penalty0 (5):\penalty0 2846--61

\bibitem[Chou et~al(2001)Chou, Kim, and Oster]{OsterBJ2001}
Chou T, Kim KS, Oster G (2001)
\newblock Statistical thermodynamics of membrane bending-mediated protein-protein attractions.
\newblock {Biophys J}, 80\penalty0 (3):\penalty0 1075--87

\bibitem[Christensen and Strange(2001)]{Strange2001}
Christensen M, Strange K (2001)
\newblock Developmental regulation of a novel outwardly rectifying mechanosensitive anion channel in {\it Caenorhabditis elegans}.
\newblock {J Biol Chem}, 276\penalty0 (48):\penalty0 45024--30

\bibitem[Clapham et~al(2001)Clapham, Runnels, and Strubing]{Clapham2001}
Clapham DE, Runnels LW, Strubing C (2001)
\newblock The TRP ion channel family.
\newblock {Nat Rev Neurosci}, 2\penalty0 (6):\penalty0 387--96

\bibitem[Cruickshank et~al(1997)Cruickshank, Minchin, Le~Dain, and
  Martinac]{Martinac1997}
Cruickshank CC, Minchin RF, Le~Dain AC, Martinac B (1997)
\newblock Estimation of the pore size of the large-conductance mechanosensitive ion channel of {\it Escherichia coli}.
\newblock {Biophys J}, 73\penalty0 (4):\penalty0 1925--31

\bibitem[Dan et~al(1993)Dan, Pincus, and Safran]{Dan1993}
Dan N, Pincus P, Safran SA (1993)
\newblock Membrane-induced interactions between inclusions.
\newblock {Langmuir}, 9:\penalty0 2768--71

\bibitem[Dan and Safran(1998)]{Dan1998}
Dan N, Safran SA (1998)
\newblock Effect of lipid characteristics on the structure of transmembrane proteins.
\newblock {Biophys J}, 75\penalty0 (3):\penalty0 1410--4

\bibitem[Doeven et~al(2005)Doeven, Folgering, Krasnikov, Geertsma, van~den
  Bogaart, and Poolman]{Doeven2005}
Doeven MK, Folgering JH, Krasnikov V, Geertsma ER, van~den Bogaart G, Poolman B (2005)
\newblock Distribution, lateral mobility and function of membrane proteins incorporated into giant unilamellar vesicles.
\newblock {Biophys J}, 88\penalty0 (2):\penalty0 1134--42

\bibitem[Duggan et~al(2000)Duggan, Garcia-Anoveros, and Corey]{Duggan2000}
Duggan A, Garcia-Anoveros J, Corey DP (2000)
\newblock Insect mechanoreception: What a long, strange TRP it's been.
\newblock {Curr Biol}, 10\penalty0 (10):\penalty0 R384--7

\bibitem[Elmore and Dougherty(2001)]{Elmore2001}
Elmore DE, Dougherty DA (2001)
\newblock Molecular dynamics simulations of wild-type and mutant forms of the {\it Mycobacterium tuberculosis} MscL channel.
\newblock {Biophys J}, 81\penalty0 (3):\penalty0 1345--59

\bibitem[Elmore and Dougherty(2003)]{Elmore2003}
Elmore DE, Dougherty DA (2003)
\newblock Investigating lipid composition effects on the mechanosensitive channel of large conductance (MscL) using molecular dynamics simulations.
\newblock {Biophys J}, 85\penalty0 (3):\penalty0 1512--24

\bibitem[Evans et~al(2003)Evans, Heinrich, Ludwig, and Rawicz]{Evans2003}
Evans E, Heinrich V, Ludwig F, Rawicz W (2003)
\newblock Dynamic tension spectroscopy and strength of biomembranes.
\newblock {Biophys J}, 85\penalty0 (4):\penalty0 2342--50

\bibitem[Fain(2003)]{Fain}
Fain GL (2003)
\newblock {Sensory Transduction}.
\newblock Sinauer Associates, Sunderland

\bibitem[Franks and Honore(2004)]{Franks2004}
Franks NP, Honore E (2004)
\newblock The TREK K2P channels and their role in general anaesthesia and neuroprotection.
\newblock {Trends Pharmacol Sci}, 25\penalty0 (11):\penalty0 601--8

\bibitem[Gambin et~al(2006)Gambin, Lopez-Esparza, Reffay, Sierecki, Gov,
  Genest, Hodges, and Urbach]{Gambin2006}
Gambin Y, Lopez-Esparza R, Reffay M, Sierecki E, Gov NS, Genest M, Hodges RS, Urbach W (2006)
\newblock Lateral mobility of proteins in liquid membranes revisited.
\newblock {Proc Natl Acad Sci U S A}, 103\penalty0 (7):\penalty0 2098--102

\bibitem[Gillespie and Walker(2001)]{Gillespie}
Gillespie PG, Walker RG (2001)
\newblock Molecular basis of mechanosensory transduction.
\newblock {Nature}, 413\penalty0 (6852):\penalty0 194--202

\bibitem[Goforth et~al(2003)Goforth, Chi, Greathouse, Providence, Koeppe, and
  Andersen]{Goforth2003}
Goforth RL, Chi AK, Greathouse DV, Providence LL, Koeppe 2nd RE, Andersen OS (2003)
\newblock Hydrophobic coupling of lipid bilayer energetics to channel function.
\newblock {J Gen Physiol}, 121\penalty0 (5):\penalty0 477--93

\bibitem[Goulian et~al(1998)Goulian, Mesquita, Fygenson, Nielsen, Andersen,
  and Libchaber]{GoulianBJ1998}
Goulian M, Mesquita ON, Fygenson DK, Nielsen C, Andersen OS, Libchaber A (1998)
\newblock Gramicidin channel kinetics under tension.
\newblock {Biophys J}, 74\penalty0 (1):\penalty0 328--37

\bibitem[Goulian et~al(1993)Goulian, Pincus, and Bruinsma]{Goulian1993}
Goulian M, Pincus P, Bruinsma R (1993)
\newblock Long-range forces in heterogenous fluid membranes.
\newblock {Europhys Letters}, 22\penalty0 (2):\penalty0 145--50

\bibitem[Gu et~al(2001)Gu, Juranka, and Morris]{MorrisBJ2001}
Gu CX, Juranka PF, Morris CE (2001)
\newblock Stretch-activation and stretch-inactivation of Shaker-IR, a voltage-gated K+ channel.
\newblock {Biophys J}, 80\penalty0 (6):\penalty0 2678--93

\bibitem[Guigas and Weiss(2006)]{Guigas2006}
Guigas G, Weiss M (2006)
\newblock Size-dependent diffusion of membrane inclusions.
\newblock {Biophys J}, 91\penalty0 (7):\penalty0 2393--8

\bibitem[Gullingsrud et~al(2001)Gullingsrud, Kosztin, and
  Schulten]{Gullingsrud2001}
Gullingsrud J, Kosztin D, Schulten K (2001)
\newblock Structural determinants of MscL gating studied by molecular dynamics simulations.
\newblock {Biophys J}, 80\penalty0 (5):\penalty0 2074--81

\bibitem[Gullingsrud and Schulten(2003)]{Gullingsrud2003}
Gullingsrud J, Schulten K (2003)
\newblock Gating of MscL studied by steered molecular dynamics.
\newblock {Biophys J}, 85\penalty0 (4):\penalty0 2087--99

\bibitem[Harroun et~al(1999)Harroun, Heller, Weiss, Yang, and
  Huang]{HuangBJ1999}
Harroun TA, Heller WT, Weiss TM, Yang L, Huang HW (1999)
\newblock Theoretical analysis of hydrophobic matching and membrane-mediated interactions in lipid bilayers containing Gramicidin.
\newblock {Biophys J}, 76\penalty0 (6):\penalty0 3176--85

\bibitem[Haswell and Meyerowitz(2006)]{Haswell}
Haswell ES, Meyerowitz EM (2006)
\newblock MscS-like proteins control plastid size and shape in {\it Arabidopsis thaliana}.
\newblock {Curr Biol}, 16\penalty0 (1):\penalty0 1--11

\bibitem[Helfrich(1973)]{Helfrich1973}
Helfrich W (1973)
\newblock Elastic properties of lipid bilayers: Theory and possible experiments.
\newblock {Z Naturforsch [C]}, 28\penalty0 (11):\penalty0 693--703

\bibitem[Hille(1968)]{Hille1968}
Hille B (1968)
\newblock Pharmacological modifications of the sodium channels of frog nerve.
\newblock {J Gen Physiol}, 51\penalty0 (2):\penalty0 199--219

\bibitem[Huang(1986)]{Huang1986}
Huang HW (1986)
\newblock Deformation free energy of bilayer membrane and its effect on Gramicidin channel lifetime.
\newblock {Biophys J}, 50\penalty0 (6):\penalty0 1061--70

\bibitem[Jensen and Mouritsen(2004)]{Jensen2004}
Jensen MO, Mouritsen OG (2004)
\newblock Lipids do influence protein function - the hydrophobic matching hypothesis revisited.
\newblock {Biochim Biophys Acta}, 1666\penalty0 (1-2):\penalty0 205--26

\bibitem[Kahya et~al(2003)Kahya, Scherfeld, Bacia, Poolman, and
  Schwille]{Kahya2003}
Kahya N, Scherfeld D, Bacia K, Poolman B, Schwille P (2003)
\newblock Probing lipid mobility of raft-exhibiting model membranes by fluorescence correlation spectroscopy.
\newblock {J Biol Chem}, 278\penalty0 (30):\penalty0 28109--15

\bibitem[Kamada et~al(1995)Kamada, Jung, Piotrowski, and Levin]{Kamada1995}
Kamada Y, Jung US, Piotrowski J, Levin DE (1995)
\newblock The protein kinase C-activated MAP kinase pathway of {\it Saccharomyces cerevisiae} mediates a novel aspect of the heat shock response.
\newblock {Genes Dev}, 9\penalty0 (13):\penalty0 1559--71

\bibitem[Katsumi et~al(2004)Katsumi, Orr, Tzima, and Schwartz]{Katsumi2004}
Katsumi A, Orr AW, Tzima E, Schwartz MA (2004)
\newblock Integrins in mechanotransduction.
\newblock {J Biol Chem}, 279\penalty0 (13):\penalty0 12001--4

\bibitem[Kloda and Martinac(2001)]{Kloda2001}
Kloda A, Martinac B (2001)
\newblock Molecular identification of a mechanosensitive channel in archaea.
\newblock {Biophys J}, 80\penalty0 (1):\penalty0 229--40

\bibitem[Lauritzen et~al(2005)Lauritzen, Chemin, Honore, Jodar, Guy,
  Lazdunski, and Jane~Patel]{Lauritzen2005}
Lauritzen I, Chemin J, Honore E, Jodar M, Guy N, Lazdunski M, Jane~Patel A (2005)
\newblock Cross-talk between the mechano-gated K2P channel TREK-1 and the actin cytoskeleton.
\newblock {EMBO Rep}, 6\penalty0 (7):\penalty0 642--8

\bibitem[Lee(2003)]{Lee2003}
Lee AG (2003)
\newblock Lipid-protein interactions in biological membranes: A structural perspective.
\newblock {Biochim Biophys Acta}, 1612\penalty0 (1):\penalty0 1--40

\bibitem[Lee(2005)]{Lee2005}
Lee AG (2005)
\newblock How lipids and proteins interact in a membrane: A molecular approach.
\newblock {Mol Biosyst}, 1\penalty0 (3):\penalty0 203--12

\bibitem[Levina et~al(1999)Levina, Totemeyer, Stokes, Louis, Jones, and
  Booth]{Levina1999}
Levina N, Totemeyer S, Stokes NR, Louis P, Jones MA, Booth IR (1999)
\newblock Protection of {\it Escherichia coli} cells against extreme turgor by activation of MscS and MscL mechanosensitive channels: Identification of genes required for MscS activity.
\newblock {Embo J}, 18\penalty0 (7):\penalty0 1730--7

\bibitem[Li et~al(2002)Li, Moe, Chandrasekaran, Booth, and
  Blount]{Blountembo2002}
Li Y, Moe PC, Chandrasekaran S, Booth IR, Blount P (2002)
\newblock Ionic regulation of MscK, a mechanosensitive channel from {\it Escherichia coli}.
\newblock {Embo J}, 21\penalty0 (20):\penalty0 5323--30

\bibitem[Maingret et~al(1999)Maingret, Fosset, Lesage, Lazdunski, and
  Honore]{Maingret1999}
Maingret F, Fosse M, Lesage F, Lazdunski M, Honore E (1999)
\newblock TRAAK is a mammalian neuronal mechano-gated K+ channel.
\newblock {J Biol Chem}, 274\penalty0 (3):\penalty0 1381--7

\bibitem[Maingret et~al(2000)Maingret, Patel, Lesage, Lazdunski, and
  Honore]{Maingret2000}
Maingret F, Patel AJ, Lesage F, Lazdunski M, Honore E (2000)
\newblock Lysophospholipids open the two-pore domain mechano-gated K+ channels TREK-1 and TRAAK.
\newblock {J Biol Chem}, 275\penalty0 (14):\penalty0 10128--33

\bibitem[Markin and Sachs(2004)]{Sachs2004}
Markin VS, Sachs F (2004)
\newblock Thermodynamics of mechanosensitivity.
\newblock {Phys Biol}, 1\penalty0 (1-2):\penalty0 110--24

\bibitem[Martinac and Hamill(2002)]{Martinac2002}
Martinac B, Hamill OP (2002)
\newblock Gramicidin A channels switch between stretch activation and stretch inactivation depending on bilayer thickness.
\newblock {Proc Natl Acad Sci U S A}, 99\penalty0 (7):\penalty0 4308--12

\bibitem[Mitra et~al(2004)Mitra, Ubarretxena-Belandia, Taguchi, Warren, and
  Engelman]{Engelman2004}
Mitra K, Ubarretxena-Belandia I, Taguchi T, Warren G,Engelman DM (2004)
\newblock Modulation of the bilayer thickness of exocytic pathway membranes by membrane proteins rather than cholesterol.
\newblock {Proc Natl Acad Sci U S A}, 101\penalty0 (12):\penalty0 4083--8

\bibitem[Moe and Blount(2005)]{Moeblount2005}
Moe P, Blount P (2005)
\newblock Assessment of potential stimuli for mechano-dependent gating of MscL:  Effects of pressure, tension, and lipid headgroups.
\newblock {Biochemistry}, 44\penalty0 (36):\penalty0 12239--44

\bibitem[Molina et~al(2006)Molina, Barrera, Fernandez, Poveda, Renart,
  Encinar, Riquelme, and Gonzalez-Ros]{Molina2006}
Molina ML, Barrera FN, Fernandez AM, Poveda JA, Renart ML, Encinar JA, Riquelme G, Gonzalez-Ros JM (2006)
\newblock Clustering and coupled gating modulate the activity in KcsA, a potassium channel model.
\newblock {J Biol Chem}, 281\penalty0 (27):\penalty0 18837--48

\bibitem[Morris and Homann(2001)]{Morris2001}
Morris CE, Homann U (2001)
\newblock Cell surface area regulation and membrane tension.
\newblock {J Membr Biol}, 179\penalty0 (2):\penalty0 79--102

\bibitem[Nauli and Zhou(2004)]{Nauli2004}
Nauli SM, Zhou J (2004)
\newblock Polycystins and mechanosensation in renal and nodal cilia.
\newblock {Bioessays}, 26\penalty0 (8):\penalty0 844--56

\bibitem[Nielsen et~al(1998)Nielsen, Goulian, and Andersen]{AndersenBJ1998}
Nielsen C, Goulian M, Andersen OS (1998)
\newblock Energetics of inclusion-induced bilayer deformations.
\newblock {Biophys J}, 74\penalty0 (4):\penalty0 1966--83

\bibitem[Niggemann et~al(1995)Niggemann, Kummrow, and Helfrich]{Helfrich1995}
Niggemann G, Kummrow M, Helfrich W (1995)
\newblock The bending rigidity of phosphatidylcholine bilayers: Dependences on experimental method, sample cell sealing and temperature.
\newblock {J Phys II France}, 5:\penalty0 413--25

\bibitem[Park and Lubensky(1996)]{ParkJFr1996}
Park JM, Lubensky TC (1996)
\newblock Interactions between membrane inclusions on fluctuating membranes.
\newblock {J Phys I France}, 6:\penalty0 1217--35

\bibitem[Perozo et~al(2002{\natexlab{a}})Perozo, Cortes, Sompornpisut, Kloda,
  and Martinac]{PerozoNature2002}
Perozo E, Cortes DM, Sompornpisut P, Kloda A, Martinac B (2002a)
\newblock Open channel structure of MscL and the gating mechanism of mechanosensitive channels.
\newblock {Nature}, 418\penalty0 (6901):\penalty0 942--8

\bibitem[Perozo et~al(2001)Perozo, Kloda, Cortes, and Martinac]{PerozoJGP2001}
Perozo E, Kloda A, Cortes DM, Martinac B (2001)
\newblock Site-directed spin-labeling analysis of reconstituted MscL in the closed state.
\newblock {J Gen Physiol}, 118\penalty0 (2):\penalty0 193--206

\bibitem[Perozo et~al(2002{\natexlab{b}})Perozo, Kloda, Cortes, and
  Martinac]{PerozoNSB2002}
Perozo E, Kloda A, Cortes DM, Martinac B (2002b)
\newblock Physical principles underlying the transduction of bilayer deformation forces during mechanosensitive channel gating.
\newblock {Nat Struct Biol}, 9\penalty0 (9):\penalty0 696--703

\bibitem[Perozo and Rees(2003)]{PerozoCOSB2003}
Perozo E, Rees DC (2003)
\newblock Structure and mechanism in prokaryotic mechanosensitive channels.
\newblock {Curr Opin Struct Biol}, 13\penalty0 (4):\penalty0 432--42

\bibitem[Pivetti et~al(2003)Pivetti, Yen, Miller, Busch, Tseng, Booth, and
  Saier]{Pivetti2003}
Pivetti CD, Yen MR, Miller S, Busch W, Tseng YH, Booth IR, Saier Jr. MH (2003)
\newblock Two families of mechanosensitive channel proteins.
\newblock {Microbiol Mol Biol Rev}, 67\penalty0 (1):\penalty0 66--85

\bibitem[Powl et~al(2003)Powl, East, and Lee]{Powl2003}
Powl AM, East JM, Lee AG (2003)
\newblock Lipid-protein interactions studied by introduction of a tryptophan residue: The mechanosensitive channel MscL.
\newblock {Biochemistry}, 42\penalty0 (48):\penalty0 14306--17

\bibitem[Rawicz et~al(2000)Rawicz, Olbrich, McIntosh, Needham, and
  Evans]{Evans2000}
Rawicz W, Olbrich KC, McIntosh T, Needham D, Evans E (2000)
\newblock Effect of chain length and unsaturation on elasticity of lipid bilayers.
\newblock {Biophys J}, 79\penalty0 (1):\penalty0 328--39

\bibitem[Sachs(1991)]{Sachs1991}
Sachs F (1991)
\newblock Mechanical transduction by membrane ion channels: A mini review.
\newblock {Mol Cell Biochem}, 104\penalty0 (1-2):\penalty0 57--60

\bibitem[Seemann and Winter(2003)]{Seemann2003}
Seemann H, Winter R (2003)
\newblock Volumetric properties, compressibilities and volume fluctuations in phospholipid-cholesterol bilayers.
\newblock {Zeitschrift fur physikalische Chemie}, 217:\penalty0 831--46

\bibitem[Shapovalov and Lester(2004)]{Lester2004}
Shapovalov G, Lester HA (2004)
\newblock Gating transitions in bacterial ion channels measured at 3 microsecond resolution.
\newblock {J Gen Physiol}, 124\penalty0 (2):\penalty0 151--61

\bibitem[Sintes and Baumgartner(1997)]{Sintes1997}
Sintes T, Baumgartner A (1997)
\newblock Protein attraction in membranes induced by lipid fluctuations.
\newblock {Biophys J}, 73\penalty0 (5):\penalty0 2251--9

\bibitem[Spencer and Rees(2002)]{Rees2002}
Spencer RH, Rees DC (2002)
\newblock The alpha-helix and the organization and gating of channels.
\newblock {Annu Rev Biophys Biomol Struct}, 31:\penalty0 207--33

\bibitem[Sukharev et~al(2001)Sukharev, Betanzos, Chiang, and
  Guy]{SukharevNature2001}
Sukharev S, Betanzos M, Chiang CS, Guy HR (2001)
\newblock The gating mechanism of the large mechanosensitive channel MscL.
\newblock {Nature}, 409\penalty0 (6821):\penalty0 720--4

\bibitem[Sukharev et~al(1997)Sukharev, Blount, Martinac, and
  Kung]{SukharevAnnRev1997}
Sukharev SI, Blount P, Martinac B, Kung C (1997)
\newblock Mechanosensitive channels of {\it Escherichia coli}: The MscL gene, protein, and activities.
\newblock {Annu Rev Physiol}, 59:\penalty0 633--57

\bibitem[Sukharev et~al(1999)Sukharev, Sigurdson, Kung, and
  Sachs]{SukharevJGP1999}
Sukharev SI, Sigurdson WJ, Kung C, Sachs F (1999)
\newblock Energetic and spatial parameters for gating of the bacterial large conductance mechanosensitive channel, MscL.
\newblock {J Gen Physiol}, 113\penalty0 (4):\penalty0 525--40

\bibitem[Tosh and Collings(1986)]{Tosh1986}
Tosh RE, Collings PJ (1986)
\newblock High pressure volumetric measurements in dipalmitoylphosphatidylcholine bilayers.
\newblock {Biochim Biophys Acta}, 859\penalty0 (1):\penalty0 10--4

\bibitem[Turner and Sens(2004)]{Turner2004}
Turner MS, Sens P (2004)
\newblock Gating-by-tilt of mechanically sensitive membrane channels.
\newblock {Phys Rev Lett}, 93\penalty0 (11):\penalty0 118103

\bibitem[Wiggins and Phillips(2004)]{Wiggins2004}
Wiggins P, Phillips R (2004)
\newblock Analytic models for mechanotransduction: Gating a mechanosensitive channel.
\newblock {Proc Natl Acad Sci U S A}, 101\penalty0 (12):\penalty0 4071--6

\bibitem[Wiggins and Phillips(2005)]{Wiggins2005}
Wiggins P, Phillips R (2005)
\newblock Membrane-protein interactions in mechanosensitive channels.
\newblock {Biophys J}, 88\penalty0 (2):\penalty0 880--902

\bibitem[Yoshimura et~al(1999)Yoshimura, Batiza, Schroeder, Blount, and
  Kung]{Yoshimura1999}
Yoshimura K, Batiza A, Schroeder M, Blount P, C Kung (1999)
\newblock Hydrophilicity of a single residue within MscL correlates with increased channel mechanosensitivity.
\newblock {Biophys J}, 77\penalty0 (4):\penalty0 1960--72

\bibitem[Yoshimura et~al(2004)Yoshimura, Nomura, and Sokabe]{Yoshimura2004}
Yoshimura K, Nomura T, Sokabe M (2004)
\newblock Loss-of-function mutations at the rim of the funnel of mechanosensitive channel MscL.
\newblock {Biophys J}, 86\penalty0 (4):\penalty0 2113--20

\end{thebibliography}

\end{document}